\documentclass[12pt, draftclsnofoot, onecolumn]{IEEEtran}

 \usepackage{amsmath}
 \usepackage{subfigure}
 \usepackage{graphicx,graphics,color,psfrag}
 \usepackage{cite,balance}
 \usepackage{amsmath,amssymb,amsfonts}
 \usepackage{caption}
 \usepackage{amsmath}
 \usepackage{tabularx}
 \usepackage{amsthm}
\usepackage{textcomp}
 \allowdisplaybreaks
 \usepackage{algorithm}
  \usepackage{algorithmic}
 \usepackage{accents}
 \usepackage{amsthm}
 \usepackage{bm}
  \usepackage{url}
  \usepackage{setspace}
 \usepackage[english]{babel}
 \usepackage{multirow}
 \usepackage{enumerate}
 \usepackage{cases}
 \usepackage{stfloats}
 \usepackage{dsfont}
 \usepackage{color,soul}
 \usepackage{amsfonts}
 \usepackage{cite,graphicx}
 \usepackage{subfigure}
 \usepackage{fancyhdr}
 \usepackage{hhline}
 \usepackage{graphicx,graphics}
 \usepackage{array,color}
 \usepackage{makecell}
 \usepackage[top=0.88in, bottom=0.88in, left=0.75in, right=0.75in]{geometry}
  \usepackage{bm}
\newtheorem{lemma}{\emph{\underline{Lemma}}}

\newtheorem{example}{\bf \emph{\underline{Example}}}
\newtheorem{remark}{\bf \emph{\underline{Remark}}}

\def\rank{{\operatorname{rank}}}

\def\l{\left}
\def\r{\right}
\def\({\left(}
\def\){\right)}

\def\b0{{\mathbf{0}}}

\newcommand{\nn}{\nonumber}

\begin{document}
\title{UAV Trajectory and Communication Co-design: Flexible Path Discretization and Path Compression}
\author{
\IEEEauthorblockN{Yijun Guo, {\it Member, IEEE}, Changsheng You, {\it Member, IEEE}, \\
Changchuan Yin, {\it Senior Member, IEEE}, and Rui Zhang, {\it Fellow, IEEE}}\\
\thanks{
Y. Guo and C. Yin are with the Beijing Key Laboratory of Network System Architecture and Convergence, Beijing University of Posts and Telecommunications, (email: \{guoyijun, ccyin\}@bupt.edu.cn).

C. You and R. Zhang  are with the Department of Electrical and Computer Engineering, National University of Singapore (email: \{eleyouc, elezhang\}@nus.edu.sg).
}}
\maketitle
\vspace{-40pt}
\begin{abstract}
The performance optimization of UAV communication systems requires the joint design of UAV trajectory and communication efficiently. 
To tackle the challenge of infinite design variables arising from the continuous-time UAV trajectory optimization, a commonly adopted approach in the existing literature  is by approximating the UAV trajectory with piecewise-linear path segments connected via a finite number of waypoints in three-dimensional (3D) space. However, this approach may still incur  prohibitive computational complexity in practice when the UAV flight period/distance becomes long, as the distance between consecutive waypoints needs to be kept sufficiently small to retain high approximation accuracy. To resolve  this fundamental  issue, we propose in this paper a \emph{new} and \emph{general}  framework 
 for UAV trajectory and communication co-design with flexible number of waypoint optimization variables (called \emph{designable} waypoints) or their \emph{sub-path} representations.  First, we propose a \emph{flexible path discretization} scheme that optimizes only a number of selected waypoints (designable waypoints) along the UAV path for complexity reduction, while all the designable and non-designable waypoints are used in calculating the approximated communication utility along the UAV trajectory for ensuring high trajectory discretization  accuracy. 
Next, given any number of designable waypoints, we  propose a novel \emph{path compression} scheme where 
  the UAV 3D path is first decomposed into three one-dimensional (1D) sub-paths and each sub-path is then   approximated by superimposing a number of  selected basis paths (which are generally less than the number of designable waypoints) weighted by their corresponding path coefficients, thus further reducing the path design complexity. Finally, we provide a case study on UAV trajectory design for aerial data harvesting from distributed ground sensors,  and numerically show that  the proposed flexible path discretization and path compression schemes can significantly reduce the UAV trajectory design complexity yet achieve favorable  rate performance as compared to conventional path/time discretization schemes.
\end{abstract}

\IEEEpeerreviewmaketitle

\section{Introduction}

Unmanned aerial vehicle (UAV) has emerged as a new promising communication platform in future wireless systems/networks, thanks to its great advantages such as controllable maneuver, high mobility, on-demand and flexible deployment, as well as line-of-sight (LoS) dominant UAV-ground channels \cite{tutor}. These appealing advantages have spurred fast-growing enthusiasm in both academia and industry recently, giving rise to a proliferation of new applications, such as  UAV-enabled relaying \cite{zeng2016throughput,chen2018local,kang20203d,chen2017optimum}, UAV-enabled data harvesting and dissemination \cite{zhan2018energy,b5,wu2018capacity,gong2018flight,you20193d,you2020hybrid},  UAV-enabled wireless power transfer \cite{xu2018uav,hu2019optimal}, cellular-connected UAV \cite{zeng2018cellular,zhang2018cellular,mei2019uplink}, among others.

Particularly, for high-mobility UAV, its  trajectory design has been extensively studied in the literature for maximizing various communication utilities (e.g., throughput, coverage, energy efficiency, and so on) under different UAV-ground channel models. For example, for UAV at high altitude in rural areas, the  LoS UAV-ground channel model is practically accurate and thus has been widely used to design the two-dimensional (2D) UAV trajectory with fixed (minimum) UAV altitude (see, e.g.,\cite{zhan2018energy,b5,wu2018capacity,gong2018flight,hua2018power,zhou2018computation}). 
While for UAV in urban areas with dense buildings, the simplified LoS UAV-ground channel model fails to capture the non-negligible multi-path fading and shadowing effects, thus two more sophisticated channel models have been proposed in the literature to improve the accuracy, namely,  the elevation-angle dependent Rician fading channel model \cite{you20193d} which is suitable for UAV operating  at high altitude, and the (generalized) probabilistic LoS channel model \cite{you2020hybrid} which is applicable to UAV at lower altitude. Under these two refined channel models, the three-dimensional (3D) UAV trajectory design has been studied for enhancing UAV communication performance by judiciously controlling the UAV-ground distance and elevation angle (both dependent on the UAV-ground horizontal and vertical distances) \cite{you20193d,you2020hybrid}. 
Note that compared to quasi-static  UAV placement design (see, e.g., \cite{mozaffari2016optimal,mozaffari2015drone,b6,kang20203d,alzenad20173}) for which only a finite number of static UAV locations need to be optimized, UAV trajectory design generally is much more challenging since it involves \emph{continuous time} that results in an \emph{infinite} number of design variables associated with UAV trajectory as well as  communication. A practical  approach to tackle this problem  is by approximating the UAV trajectory with a  tractable \emph{piecewise-linear continuous} trajectory where the path comprises consecutive line segments connected via a \emph{finite} number of waypoints in 3D, while  the time duration that the UAV spends on each line segment can be different. This scheme  is referred to as the  \emph{conventional path discretization} (CPD) in this paper. Besides,  another widely-used trajectory discretization scheme, called \emph{time discretization} (TD), can be regarded as a special case of CPD with \emph{identical} time-slot length over all line segments.
    To ensure sufficiently high trajectory  discretization accuracy for the UAV-ground communication performance, it is required that for both CPD and TD schemes, their adopted segment lengths should be no larger than a certain threshold such that the UAV-ground distance can be regarded as approximately unchanged within each line segment.

 The above two trajectory discretization schemes, however, may entail a large number of line segments in practice when the UAV flight distance/period becomes long, thus resulting in prohibitive computational complexity for the 2D/3D UAV trajectory design. Moreover, to address the coupling between UAV trajectory and communication designs, a commonly adopt approach in the existing literature is by using the block coordinate descent (BCD) method to decouple the joint optimization via alternately optimizing UAV trajectory and communication in an iterative manner, which incurs additional computational complexity. 
 These issues have motivated active research efforts recently to reduce the UAV trajectory design complexity \cite{zhang2019receding,lee2019uav,guo2020novel,hu2019optimal,you2020hybrid,shen2020multi,xu2020low}. For example, a low-complexity receding-horizon  based optimization method was proposed in \cite{zhang2019receding} that progressively designs the overall UAV trajectory with a moving time-window. Specifically, different from TD that imposes equal time-slot length on all line segments, the authors proposed to divide the entire time horizon into three sub-horizons with different time-slot lengths so as to reduce the number of design variables in each time-window. 
Alternatively, to maximize the UAV flight time given fixed propulsion energy supply, a customized low-complexity UAV trajectory design was proposed in \cite{lee2019uav} that  optimizes only a UAV trajectory fragment in a short time horizon, while it treats the overall trajectory as the rotated and replicated ones of the trajectory fragment with certain rotation angles.  In \cite{guo2020novel}, the authors proposed to fit the UAV trajectory over time by finite Fourier series with TD, based on which  the complicated UAV waypoint optimization can be reduced into finding optimal values of finite Fourier series coefficients.	
	In \cite{hu2019optimal}, the authors considered the one-dimensional (1D) UAV trajectory and transformed the waypoint  optimization problem into optimizing UAV hovering locations and durations by leveraging its optimal hover-and-fly structure.
	Besides, a hybrid offline-online optimization framework was developed in \cite{you2020hybrid} to reduce the real-time trajectory design complexity, which first solves a time-consuming path optimization problem in the offline phase and then refines  the UAV flying speeds and communication scheduling in the online phase by solving a low-complexity linear programming (LP).
 On the other hand, to reduce the complexity arising from the BCD method,  an initial attempt has been made in \cite{shen2020multi} that simultaneously optimizes UAV trajectory and transmit power allocation.
In addition, the alternating direction method of multipliers (ADMM) method was applied in \cite{xu2020low} to decouple the joint optimization and obtain closed-form solutions to corresponding subproblems for reducing the computational complexity. 
Nevertheless, in view of the above works, there still lacks a \emph{general} framework to reduce the complexity of UAV trajectory and communication co-design from the perspective of trajectory discretization/compression, which thus  motivates our current work as the first attempt to fill this gap, to the authors' best knowledge.

To this end, we  first formulate a generic optimization problem to maximize the UAV communication utility by jointly designing the continuous-time UAV trajectory and communication. We then consider the piecewise-linear UAV trajectory for tractability and revisit the existing CPD and TD schemes as well as discuss their practical implementation issues.
Next, to reduce the computational complexity of the CPD and TD schemes with practically large number of line segments, we propose a new and general framework that is able to reduce the number of (trajectory and communication) design variables in both the \emph{time} and \emph{spatial} domains. The main contributions of this paper are summarized as follows.

\begin{itemize}
\item
Firstly,  we propose a novel \emph{flexible path discretization} (FPD) scheme to reduce the  number of waypoints that need to be optimized over time in the CPD scheme. Specifically, we divide the waypoints required  for ensuring desired trajectory discretization accuracy into two exclusive  sets, namely, \emph{designable} and \emph{non-designable}  waypoints. Among them, only the  designable waypoints are optimized for reducing the UAV trajectory design complexity, while all the waypoints (including both designable and non-designable ones) are used in calculating the approximated utility and constraint functions to retain high trajectory discretization accuracy. Moreover, we  show that the proposed FPD scheme in general entails fewer (designable) waypoints for trajectory representation as compared to TD and CPD schemes.
\item Secondly, given any number of (designable) waypoints, we propose another alternative method, called \emph{path compression} (PC), to further reduce the number of UAV trajectory design variables in the spatial domain. To this end, we first show that the UAV 3D path can be equivalently decomposed into three 1D sub-paths in the three coordinates, respectively,  each of which can be represented by  a superposition  of the same number of basis paths as that of the time-domain waypoints,  weighted by their corresponding path coefficients. Based on this result, we then  propose  a simple yet efficient PC method to approximate each sub-path with a reduced number of properly selected basis paths and their corresponding designable path coefficients, thus further reducing the design complexity of the FPD/CPD/TD schemes.  
\item Finally, we provide a case study to show different formulations of the same trajectory and communication co-design problem for the application of  UAV-enabled data harvesting from distributed sensors under different trajectory discretization schemes with/without PC. By analyzing their computational complexities for solving their respective optimization problems, we show that compared to TD and CPD schemes, the UAV trajectory and communication co-design with the proposed FPD scheme achieves much lower complexity, which can be further reduced when combined with the proposed PC scheme. Moreover,  the proposed FPD and PC schemes also achieve favorable max-min rate performance as compared to TD and CPD schemes.
\end{itemize}

The remainder of this paper is organized as follows. Section~\ref{Problem} introduces the generic problem formulation and existing trajectory discretization schemes. Then,  the proposed FPD and PC schemes are presented in Sections~\ref{SecFPD} and \ref{SecCompre}, respectively. A case study of UAV trajectory design for aerial data harvesting from distributed sensors is provided in Section~\ref{Case},  followed by the corresponding simulation results given in Section~\ref{Simulation}. Finally, the conclusions are drawn  in Section~\ref{SecConclusion}. For ease of reference, we summarize in  Table~\ref{Symbol} the main symbols and abbreviations used in this paper.


\begin{table}[t]\small
\caption{List of main symbols/abbreviations  and their meanings.}\label{Symbol}
\centering
\begin{spacing}{1.15}
\begin{tabular}{|c|l|}
\hline
$I$ 	&	Number of constraints\\\hline
$N$		&	Number of segments by conventional path discretization\\\hline
$M$ 	&	Number of segments by time discretization\\\hline
$L$ 	&	Number of long-segments by flexible path discretization\\\hline
$J$ 	&	Number of short-segments within each long-segment by flexible path discretization\\\hline
$N_{\rm FPD}$		&	Number of all short-segments by flexible path discretization\\\hline
$K$	&	Number of selected basis paths for path compression\\\hline
$U$, $\overline{U}$  & Communication utility function and its finite-sum approximation\\\hline
$f$, $\overline{f}$  & Constraint function and its finite-sum approximation\\\hline
TD  & Time discretization\\\hline
CPD  & Conventional path discretization\\\hline
FPD & Flexible path discretization\\\hline
PC  & Path compression\\\hline
FPD-PC  & Flexible path discretization with path compression\\
\hline
\end{tabular}
\end{spacing}
\end{table}

\section{Problem Formulation and Trajectory Discretization}\label{Problem}

\subsection{Generic Problem Formulation}
The performance optimization of UAV communication systems in general  requires the joint design of UAV trajectory and communication. 
Without loss of generality, let $\mathcal{Q}(t)$ denote the UAV trajectories of all UAVs over time $t\in[0,T]$ with $T$ denoting the total UAV flight period,  and $\mathcal{R}(t)$ represent all relevant variables associated with the communication design over time $t$, such as communication scheduling, bandwidth and transmit power allocation, beamforming, etc. Then, a generic optimization problem for maximizing the UAV communication utility via jointly designing UAV trajectory and communication can be formulated as follows \cite{tutor}.
\begin{subequations}
\begin{align} 
	{{\rm (P1)}:} ~~\mathop{\max}_{\mathcal{Q}(t), \mathcal{R}(t)}   \quad & U(\mathcal{Q}(t), \mathcal{R}(t))\nn \\
	{\rm s.t.} \!\!\qquad
	&f_i(\mathcal{Q}(t))\le 0, ~\qquad\quad i=1,\cdots,I_1,\label{Eq:P1traj}\\
	&h_i(\mathcal{R}(t))\le 0, ~\qquad\quad i=1,\cdots,I_2,\label{Eq:P1com}\\
	&g_i(\mathcal{Q}(t),\mathcal{R}(t))\le 0, \quad i=1,\cdots,I_3,\label{Eq:P1trajcom}
\end{align} 	
\end{subequations}
where $U(\mathcal{Q}(t), \mathcal{R}(t))$ denotes the communication utility function (such as communication throughput, energy efficiency, etc.) with respect to (w.r.t.) both UAV trajectory and communication variables in general; $f_i(\mathcal{Q}(t)), i=1,\cdots,I_1$ represent the set of constraints on  UAV trajectory only (e.g., maximum UAV speed); $h_i(\mathcal{R}(t)), i=1,\cdots,I_2$ specify the set of constraints on UAV communication only; and $g_i(\mathcal{Q}(t),\mathcal{R}(t)), i=1,\cdots,I_3$ denote the set of coupled constraints related to both UAV trajectory and communication. 
Problem ($\rm P 1$), in general, is difficult to be efficiently and optimally solved due to the following two main  reasons. First, it  involves continuous time $t$ and thus results in  an infinite number of UAV trajectory and communication design variables. Second, the UAV trajectory and communication design variables need to be jointly optimized, which usually renders problem ($\rm P 1$) a non-convex optimization problem and thus hard to be optimally solved. 

To tackle the above difficulties, we can first apply the BCD method to decouple the joint optimization for UAV trajectory and communication. Specifically, by regarding $\mathcal{Q}(t)$ and $\mathcal{R}(t)$ as two blocks of design variables,  the non-convex problem ($\rm P 1$) can be sub-optimally  solved  by using  an iterative algorithm as follows. On one hand, given any feasible $\mathcal{Q}(t)\!=\!\hat{\mathcal{Q}}(t)$, problem ($\rm P1$) is equivalent to
\begin{subequations}
\begin{align} 
	\!\!\!\!\!{{\rm (P2)}:} ~\mathop{\max}_{\mathcal{R}(t)}   \quad & U(\hat{\mathcal{Q}}(t), {\mathcal{R}}(t))\nn \\
	{\rm s.t.} \quad
	&h_i(\mathcal{R}(t))\le 0, ~\qquad\qquad i=1,\cdots,I_2,\label{Eq:P1.1com}\\
	&g_i(\hat{\mathcal{Q}}(t),{\mathcal{R}}(t))\le 0, \qquad i=1,\cdots,I_3,\label{Eq:P1.1trajcom}
\end{align} 
\end{subequations}
which reduces to  the communication design problem only.  By discretizing the time into a finite number of intervals (e.g., TD  introduced later), problem ($\rm P2$) can be efficiently solved by using existing techniques developed for wireless communication resource-allocation optimization over parallel channels (e.g., block-fading channels, orthogonal-frequency channels). On the other hand, given any feasible $\mathcal{R}(t)=\hat{\mathcal{R}}(t)$, problem ($\rm P1$) reduces to
\begin{subequations}
\begin{align} 
	\!\!\!\!\!{{\rm (P3)}:} ~\mathop{\max}_{\mathcal{Q}(t)}   \quad & U(\mathcal{Q}(t), \hat{\mathcal{R}}(t))\nn \\
	{\rm s.t.} \quad
	&f_i(\mathcal{Q}(t))\le 0, ~\qquad\qquad i=1,\cdots,I_1,\label{Eq:P1.2traj}\\
	&g_i(\mathcal{Q}(t),\hat{\mathcal{R}}(t))\le 0, \qquad i=1,\cdots,I_3,\label{Eq:P1.2trajcom}
\end{align} 
\end{subequations}
which needs to optimize the \emph{continuous-time} UAV trajectory. 
For the purpose of exposition,  we focus on  the case with one single UAV only in this paper, where $\mathcal{Q}(t)\triangleq{\bf q}(t)$ with ${\bf q}(t)$ denoting the  single-UAV 3D trajectory over time, while the proposed design framework and solutions can be extended to the general case with multiple UAVs (e.g., a swarm of UAVs). By slightly abusing the notations of $U(\cdot)$ and $f(\cdot)$, problem ($\rm P3$) 
 can be equivalently rewritten as
 \vspace{-5pt} 
\begin{subequations}
\begin{align} 
	\!\!\!\!\!{{\rm (P4)}:} ~~\mathop{\max}_{{\bf q}(t)}   \quad & U({\bf q}(t))\nn \\
	{\rm s.t.} \quad
	&f_i({\bf q}(t))\le 0, ~\qquad\qquad i=1,\cdots,I,\label{Eq:P4traj}
\end{align} 
\end{subequations}
where $I=I_1+I_3$, and $f_i({\bf q}(t)), i=1,\cdots,I$ represent the set of constraints in \eqref{Eq:P1.2traj} and \eqref{Eq:P1.2trajcom}  associated with UAV trajectory. Compared to problem ($\rm P2$), problem ($\rm P4$) is relatively new as well as more challenging to solve  due to the continuous-time trajectory and trajectory-dependent communication utility/constraints. As such, we focus on solving problem ($\rm P4$)  in the rest of this paper.

\vspace{-6pt}
\subsection{UAV Trajectory Discretization:  Existing Schemes}\label{SecPath}
\begin{figure}[t] \centering    
	\includegraphics[width=0.9\columnwidth]{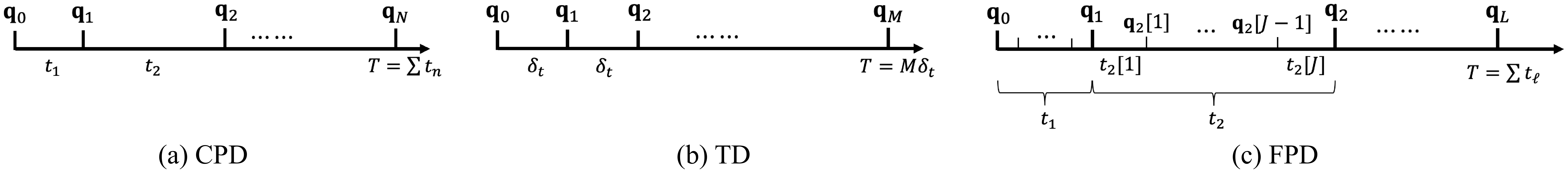} 
	\caption{Illustration of different piecewise-linear trajectory discretization schemes in 1D.}     
	\label{TDPDFPD}
	\end{figure}
	
\subsubsection{CPD}

The solution to problem ($\rm P4$), in general,  is intractable due to the infinite number of UAV trajectory variables over time, i.e., ${\bf q}(t)$. To overcome this difficulty, we consider the tractable \emph{piecewise-linear} continuous UAV trajectory as illustrated in Fig.~\ref{TDPDFPD}(a), which can be  characterized by a sequence of \emph{ordered waypoints} along the UAV flying \emph{path} with line segments connecting them, as well as the \emph{traveling time duration} that the UAV spends on each line segment. Thereby, 
the UAV velocity (specifying both direction and speed) over each line segment is also determined if it  is assumed to be constant over each line segment (but can change over different segments).
 Mathematically, 
let $\{{\bf q}_n\}_{n=0}^{N}$ denote the $N+1$ (ordered) waypoints along the UAV path with ${\bf q}_n\triangleq[q_{n, {\rm x}}, q_{n, {\rm y}}, q_{n, {\rm z}}]^T$. For each of the resultant  $N$ line segments, say, segment $n$ that connects the waypoints ${\bf q}_{n-1}$ and ${\bf q}_n$, we denote by $t_n$ the UAV traveling duration over it. As such, the piecewise-linear continuous trajectory ${\bf q}(t)$ can be fully characterized by $\{\{{\bf q}_n\}_{n=0}^{N}, \{t_n\}_{n=1}^{N}\}$, based on which, the constant  UAV velocity over each segment $n$, denoted by ${\bf v}_n$, is given by $\mathbf{v}_n\triangleq (\mathbf{q}_n\!-\!\mathbf{q}_{n-1})/t_n,  n=1, \cdots, N$. It is worth mentioning that the above piecewise-linear trajectory is also called \emph{path discretization} in the existing literature \cite{tutor}, as the UAV path is discretized into multiple line segments with flexible lengths and UAV traveling durations over them. We refer to this scheme as CPD in this paper for convenience.  

Although much simplified, the functions of communication utility and constraints under the piecewise-linear continuous trajectory,  i.e., $U({\bf q}(t))$  and $\{f_i({\bf q}(t))\}$, are still hard to obtain in closed-form  for optimization in general, since they involve integrals w.r.t. the continuous UAV trajectory over each segment. 
An  efficient approach to tackle this difficulty is via approximating the integral by a finite-sum. Taking  the communication utility function as an example, its finite-sum approximation can be obtained as
\vspace{-10pt}
\begin{align}
U({\bf q}(t))&=\int_0^T u({\bf q}(t))  dt\approx \sum_{n=1}^{N} t_n u({\bf q}(\tau_n))= \sum_{n=1}^{N} t_n u({\bf q}_n)\triangleq \bar{U}_{{\rm CPD}}(\{{\bf q}_n\},\{t_n\}),\label{Eq:FiniteSumAppro}
\end{align}
where $u(\cdot)$ denotes the utility function associated with UAV locations, ${\bf q}(\tau_n)$ denotes any UAV location over segment $n$ (with $\sum_{\tilde{n}=1}^{n-1} t_{\tilde{n}}\le \tau_n \le \sum_{\tilde{n}=1}^{n} t_{\tilde{n}}$) which is set as ${\bf q}(\tau_n)={\bf q}(\sum_{\tilde{n}=1}^{n} t_{\tilde{n}})\triangleq{\bf q}_n$ without loss of generality, and $\bar{U}_{{\rm CPD}}(\cdot)$ represents the approximated communication utility function with CPD.   
To ensure sufficiently high finite-sum approximation accuracy, the length of each line segment, intuitively, should not exceed a certain threshold. To show this, we first
  present a useful lemma below that characterizes the finite-sum approximation error for the utility function given in \eqref{Eq:FiniteSumAppro}.
\begin{lemma}\label{lemErrorbound}\emph{Let $D_u=\mathop{\max}_{{\mathbf{q}}\in{\bf q}(t)}\|\nabla u({\mathbf{q}})\|$ denote the maximum norm of the gradient of the utility function $u({\bf q}(t))$  given any feasible trajectory ${\bf q}(t)$, and assume that $D_u$ exists and $D_u<\infty$. Then, given $T$ and the maximum  distance between consecutive  waypoints for approximating the utility function given in \eqref{Eq:FiniteSumAppro}, denoted by $\Delta_{\max}^{U}$,  the finite-sum approximation  error is upper-bounded  by
\begin{align}
E_U=| U({\bf q}(t))- \bar{U}_{{\rm CPD}}(\{{\bf q}_n\},\{t_n\})|\le\frac{1}{2} D_u \Delta_{\rm max}^{{U}} T.
\end{align}
}
\end{lemma}
\begin{proof}
See Appendix~\ref{Errorbound}.
\end{proof}
With a prescribed maximum tolerable utility approximation error $E_{U,\max}$,  $\Delta_{\max}^{{U}}$ 
should be set as 
\vspace{-6pt}
\begin{align}\label{utility_continuous}
	\Delta_{\max}^{{U}}=\frac{2 E_{U,\max}}{T D_{u}}.
\end{align}
\begin{example}\label{Example1}\emph{Consider a UAV-enabled data harvesting system, where a UAV is dispatched to collect data from one ground sensor node (SN) located at ${\bf w}\in\mathbb{R}^{3\times 1}$. Assuming the simplified 
LoS UAV-ground channel model, 
 the communication utility, defined as the average achievable rate in bits per second per Hertz (bps/Hz) at the UAV within a flight period of $T$, is given by \cite{b5}
 \vspace{-8pt}
\begin{align}\label{utility_continuous}
	U(\mathbf{q}(t))=\frac{1}{T}\int_0^T\log_2\left( 1+ \frac{P\beta_0}{\|\mathbf{q}(t)-\mathbf{w}\|^2\sigma^2} \right) dt,
\end{align}
where $P$ denotes the SN's transmit power, $\beta_0$ denotes the channel power gain at the reference  distance of $1$ meter (m), and $\sigma^2$ is the receiver noise power at the UAV. With CPD and based on \eqref{Eq:FiniteSumAppro}, the  utility function in \eqref{utility_continuous}
 can be approximated by
\begin{align}
U(\mathbf{q}(t))&\approx \bar{U}_{\rm CPD}(\{{\bf q}_n\},\{t_n\})= \frac{1}{T}\sum_{n=1}^{N} t_n \log_2 \left(1+\frac{P\beta_0}{\|\mathbf{q}_n-\mathbf{w}\|^2\sigma^2}\right).\label{Eq:PDutility}
\end{align}
Then, as proved in Appendix~\ref{rateBound}, for $u({\bf q}(t))=\frac{1}{T}\log_2 \left(1+\frac{P\beta_0}{\|\mathbf{q}(t)-\mathbf{w}\|^2\sigma^2}\right)$ and given the  minimum  UAV  flight altitude $H_{\min}$, we have $D_u=\frac{2c_2}{\ln2}\frac{c_1}{(c_1^2+H_{\min}^2)(c_1^2+H_{\min}^2+c_2)}$ where  the constants $c_1$ and $c_2$ are  defined in Appendix~\ref{rateBound}. As such, given the maximum tolerable utility approximation error $E_{U,\max}$, the segment length should satisfy $\|\mathbf{q}_n-\mathbf{q}_{n-1}\|\le\Delta_{\rm max}^{{U}}= \frac{2 E_{U,\max}}{T D_{u} }, n=1,\cdots, N$. Moreover, it can be shown that $\Delta_{\rm max}^{{U}}$ monotonically increases with $H_{\min}$, which is expected since the UAV-ground distance changes more slowly when the UAV is at higher altitude. 
}
\end{example}
By using the similar method as above, we can also  obtain the maximum segment length associated with each constraint, which is denoted by  $\Delta_{\max}^{{f}_i}, i=1,\cdots, I$. 
With $\Delta_{\max}^{{U}}$ and $\{\Delta_{\max}^{{f}_i}\}$, the maximum segment length that satisfies all the finite-sum approximation requirements can be obtained as  $\Delta_{\max}=\min\{\Delta_{\max}^{{U}}, \{\Delta_{\max}^{{f}_i}\}\}$. As such, we have 
\vspace{-6pt}
\begin{align}
\|\mathbf{q}_n-\mathbf{q}_{n-1}\| \leq \min\{\Delta_{\max}, t_nV_{\max}\}, ~~n=1,\cdots, N,\label{Eq:PTdis}
\end{align}
which incorporates  the UAV maximum  speed constraint under the CPD scheme with $V_{\max}$ denoting the maximum UAV speed.
Let $\mathbf{q}_{\rm str}$ and $\mathbf{q}_{\rm end}$ denote the start and end locations for the UAV trajectory (if specified). Then, if $\mathbf{q}_{\rm str}\neq\mathbf{q}_{\rm end}$,  the  number of line segments, $N$, should satisfy $N\ge N_{\min}\triangleq\l\lceil\frac{\|{\bf q}_{\rm str}-{\bf q}_{\rm end}\|}{\Delta_{\max}}\r\rceil$,
where $\lceil \cdot \rceil$ denotes the ceiling function and the equality holds when the UAV flies toward the end location following a straight-line path. Note that given  $T\ge \frac{\|\mathbf{q}_{\rm str}-\mathbf{q}_{\rm end}\|}{V_{\max}}$, the UAV trajectory design becomes more flexible as $T$ increases. Hence, we usually  set $N\gg N_{\min}$ when $T$ is practically  large.

Under the above finite-sum approximation for both the communication utility $U({\bf q}(t))$  and constraint functions $\{f_i({\bf q}(t))\}$, 
problem ($\rm P4$) can be approximately reformulated as
\vspace{-6pt}
\begin{subequations}
\begin{align} 
	\!\!\!\!\!{{\rm (P5)}:} ~\mathop{\max}_{\{{\bf q}_n\},\{t_n\}}   \quad & \bar{U}_{\rm CPD}(\{{\bf q}_n\},\{t_n\})\nn \\
	{\rm s.t.}~~ \quad
	&\bar{f}_{i,{\rm CPD}}(\{{\bf q}_n\},\{t_n\})\le 0, ~ i=1,\cdots,I,\label{Eq:P2traj}
\end{align} 
\end{subequations}
where $\{\bar{f}_{i,{\rm CPD}}(\{{\bf q}_n\},\{t_n\})\}$ denote the set of approximated constraints under the CPD scheme. Although problem ($\rm P5$)  is generally  non-convex,  the BCD and successive convex approximation (SCA) techniques  (as will be detailed in Section~\ref{Case}) can be employed to solve it sub-optimally.

\begin{remark}[Practical implementation]\emph{Let $\{{\bf q}_n^*\}$, and $\{t_n^*\}$ denote the optimized UAV waypoints and time durations by solving problem ($\rm P5$), respectively, and ${\bf q}^*(t)$ denote the piecewise-linear trajectory constructed from $\{\{{\bf q}_n^*\},\{t_n^*\}\}$.  It is worth noting that  the attainable communication utility by following the designed trajectory ${\bf q}^*(t)$ (i.e., $U({\bf q}^*(t))$) may not be the same as the estimated one (i.e., $\bar{U}_{\rm CPD}(\{{\bf q}_n^*\},\{t_n^*\})$) due to the finite-sum approximation error, which, however, can be upper-bounded if the conditions in Lemma~\ref{lemErrorbound} are satisfied. Moreover, the constraints of $f_i({\bf q}^*(t))\le 0,  i=1,\cdots,I$ may also be violated in practice. To address this issue, an efficient approach is to \emph{proactively} introduce robustness to the constraints prior to  optimization so as to reduce the constraint-violation probability in practical implementation. For instance, we can redefine the constraints as  
\begin{align}
(1+\epsilon_i)\bar{f}_{i,{\rm CPD}}(\{{\bf q}_n\},\{t_n\})\le 0, ~ i=1,\cdots,I,
\end{align}
by setting the parameter $\epsilon_i$ according to the prescribed finite-sum approximation accuracy requirement.
}
\end{remark}
\subsubsection{TD}
As illustrated  in Fig.~\ref{TDPDFPD}(b), TD is a special case of CPD, for which the given time horizon $[0,T]$ is divided into $M$ equal-time slots with sufficiently short slot length $t_m=\delta_t\triangleq T/M , m=1, \cdots, M$. Similarly, we can  apply the finite-sum approximation to the utility  and constraint functions  given the same finite-sum approximation accuracy as CPD. For example, the communication utility function can be approximated as
\vspace{-8pt}
\begin{align}
U({\bf q}(t))\approx \sum_{m=1}^{M} t_m u({\bf q}_m)=\delta_t  \sum_{m=1}^{M} u({\bf q}_m)\triangleq \bar{U}_{\rm TD}(\{{\bf q}_m\}).
\end{align} 
Since each segment length should not exceed $\Delta_{\max}$ for ensuring the desired finite-sum approximation accuracy even at the maximum UAV speed,
we have $\delta_t\le \frac{\Delta_{\max}}{V_{\max}}$. Thus, the number of time slots for TD, $M$, should satisfy $M=\frac{T}{\delta_t}\ge\frac{TV_{\max}}{\Delta_{\max}}$.  
Based on the above, with TD and the finite-sum approximation, 
problem ($\rm P4$) can be reformulated as
\vspace{-6pt}
\begin{subequations}
\begin{align} 
	\!\!\!\!\!{{\rm (P6)}:} ~\mathop{\max}_{\{{\bf q}_m\}}   \quad & \bar{U}_{\rm TD}(\{{\bf q}_m\})\nn \\
	{\rm s.t.} \quad
	&\bar{f}_{i,{\rm TD}}(\{{\bf q}_m\})\le 0, ~ i=1,\cdots,I,\label{Eq:P5traj}
\end{align} 
\end{subequations}
where $\{\bar{f}_{i,{\rm TD}}(\{{\bf q}_m\})\}$ denote the set of approximated constraints with TD.

Recall that for the CPD scheme, the time duration over each line segment can be flexibly adjusted under the constraint of  $t_n \le \frac{\Delta_{\max}}{V_{n}}$ where $V_{n}= \|\mathbf{v}_n\|$. Comparing it with the time-slot length of TD (i.e., $\delta_t \le \frac{\Delta_{\max}}{V_{\max}}$), we have $\delta_t\le t_n$ since $V_{n}\le V_{\max}$,  which indicates that given the same  $T$ and $\Delta_{\max}$, TD in general entails shorter time-slot length than CPD. Combining this with $\sum_{n=1}^N t_n = M\delta_t=T$, we conclude  that TD usually requires more line segments than CPD (i.e., $M\ge N$).   As shown in \cite{tutor}, the computational complexity for UAV trajectory and communication co-design mainly lies in the waypoint optimization. Generally speaking, the more waypoints that need  to be optimized, the higher is the UAV trajectory design complexity. As such, the UAV trajectory design with CPD usually incurs less computational complexity as compared to that with TD.

\vspace{-8pt}
\section{Proposed Flexible Path Discretization}\label{SecFPD}
In this section,  we propose a new trajectory discretization scheme, called FPD, to reduce the UAV trajectory design complexity as compared to the existing CPD scheme in \cite{tutor}.

\subsection{Flexible Path Discretization}	
Note that for the CPD scheme, the number of waypoints $\{{\bf q}_n\}$ that need to be optimized may be practically large, when there is a stringent finite-sum approximation accuracy requirement (i.e., small $\Delta_{\max}$) and/or the flight distance becomes long.
In this case, optimizing all the $N+1$ waypoints for maximizing the communication utility may incur prohibitive computational complexity. To resolve this issue, a key observation is that optimizing \emph{part} of the $N+1$ waypoints  may not affect the finite-sum approximation accuracy (i.e., the maximum approximation error), provided that all the $N+1$ waypoints are employed in calculating the finite-sum approximation for the utility  and constraint functions.
 Motivated by this, we propose a novel FPD scheme that divides the $N+1$ waypoints (for ensuring high finite-sum approximation accuracy) into two exclusive  sets, namely, \emph{designable} and \emph{non-designable} waypoints, which are constructed as follows and illustrated in Fig.~\ref{TDPDFPD}(c).
\begin{itemize}
\item[1)]\textbf{Designable waypoints:} First, we divide the UAV path into $L$ consecutive \emph{long-(line)-segments} of generally unequal lengths with $L\le N$. The $L+1$ waypoints connecting these long-segments, denoted by $\{{\bf q}_{\ell}\}_{\ell=0}^{L}$, are referred to as the {designable waypoints} which are used in both the (utility and constraints) function finite-sum approximation and UAV trajectory optimization. For each long-segment $\ell\in\{1, \cdots, L\}$, we denote by $t_{\ell}$ the UAV traveling  time  over it. 
\item[2)]\textbf{Non-designable waypoints:} Next, given the $L+1$ designable waypoints $\{{\bf q}_{\ell}\}_{\ell=0}^{L}$, we further divide each long-segment, say, $\ell$, into $J$ \emph{short-(line)-segments} of \emph{equal} length, which are characterized by the  two consecutive  designable waypoints at its two ends, ${\bf q}_{\ell-1}$ and ${\bf q}_{\ell}$, as well as $J-1$ waypoints along the long-segment between them. These  $J-1$ waypoints are thus called \emph{non-designable} waypoints that are not involved in UAV trajectory optimization for reducing the design complexity, but still used in the finite-sum approximation for the utility and constraints for retaining the desired accuracy. Mathematically, given ${\bf q}_{\ell-1}$ and ${\bf q}_{\ell}$, the $J-1$ non-designable  waypoints over each long-segment $\ell$, denoted by $\{\mathbf{q}_{\ell}[j]\}_{j=1}^{J-1}$, can be obtained as
\begin{align} 
	\!\!\!\!\!\!\mathbf{q}_{\ell}[j]&={\bf q}_{\ell-1}+\frac{j({\bf q}_{\ell}-{\bf q}_{\ell-1})}{J},~~ j=1, \cdots, J-1.
	\end{align}
By assuming constant UAV speed over each short-segment,  the time duration for the $j$-th short-segment of long-segment $\ell$, denoted by $t_{\ell}[j]$, is given by $t_{\ell}[j]=\frac{t_{\ell}}{J}, j=1, \cdots, J$.
\end{itemize}

Based on the above, the UAV trajectory under the  proposed FPD scheme is represented by $N_{\rm FPD}\triangleq LJ$ consecutive short-segments connected by $L+1$ designable waypoints and $L(J-1)$ non-designable  waypoints, as well as the time duration that the UAV spends on each of the $L$  long-segments (or equivalently, each of its short-segments with given $J$). By using the same method as in \eqref{Eq:FiniteSumAppro}, the communication utility with the finite-sum approximation applied to  the $N_{\rm FPD}$ short-segments is given by 
\vspace{-10pt}
\begin{align}
\!\!\!\!U({\bf q}(t))&\approx \sum_{\ell=1}^{L}\l[\l(\sum_{j=1}^{J-1} t_{\ell}[j] u(\mathbf{q}_{\ell}[j]) \r)+ t_{\ell}[J] u(\mathbf{q}_{\ell})\r]\nn\\
&=\sum_{\ell=1}^{L}\l[ \frac{t_{\ell}}{J} \sum_{j=1}^{J}  u\l({\bf q}_{\ell-1}+\frac{j({\bf q}_{\ell}-{\bf q}_{\ell-1})}{J}\r) \r]\triangleq \bar{U}_{{\rm FPD}}(\{{\bf q}_{\ell}\},\{t_{\ell}\}),\label{Eq:FleSumAppro}
\end{align}
which is determined by the designable waypoints and long-segment time durations only. 
Comparing \eqref{Eq:FleSumAppro} and \eqref{Eq:FiniteSumAppro}, we observe  that given the same finite-sum approximation accuracy as CPD, the short-segment length should not exceed the threshold $\Delta_{\max}$ and hence the long-segment length should satisfy 
\begin{align} 
	\|{\bf q}_{\ell}-{\bf q}_{\ell-1}\| \leq J\Delta_{\rm max}, ~~\ell=1,\cdots,L.
\end{align}
As such, with the proposed FPD and finite-sum approximation, problem ($\rm P4$) can be transformed into the following approximated form.
 \begin{subequations}
\begin{align} 
	\!\!\!\!\!{{\rm (P7)}:} ~\mathop{\max}_{\{{\bf q}_{\ell}\},\{t_{\ell}\}}   \quad & \bar{U}_{\rm FPD}(\{{\bf q}_{\ell}\},\{t_{\ell}\})\nn \\
	{\rm s.t.}~~ \quad
	&\bar{f}_{i,{\rm FPD}}(\{{\bf q}_{\ell}\},\{t_{\ell}\})\le 0, ~ i=1,\cdots,I,\label{Eq:P2traj}
\end{align} 
\end{subequations}
where $\{\bar{f}_{i,{\rm FPD}}(\{{\bf q}_{\ell}\},\{t_{\ell}\})\}$ denote the set of approximated constraints under the FPD scheme. 

Note that different from CPD for which all the $N+1$ waypoints required for ensuring desired finite-sum approximation accuracy are optimized for maximizing the communication utility, the proposed FPD scheme optimizes only part of the $N+1$ waypoints for reducing UAV trajectory design  complexity, while all the $N_{\rm FPD}+1=N+1$ (designable and non-designable) waypoints are employed in calculating the finite-sum approximation for the utility and constraints to satisfy the approximation accuracy requirements.  Moreover, for the proposed FPD scheme, there exists a fundamental trade-off between UAV trajectory design complexity and communication utility performance via adjusting the number of designable waypoints, $L+1$. Specifically, as $L$ decreases, the UAV trajectory design complexity reduces due to the smaller number of designable waypoints, while the design degrees-of-freedom (DoF) for UAV trajectory also degrades  since the UAV can change its flying direction and speed along the path for at most $L-1$ times. As such,  small $L$ may result in less optimal trajectory and thus certain utility loss as compared to that with larger $L$.  
In the special case with $L=N$, the proposed FPD reduces to CPD.

\subsection{Flexible Path Discretization versus Path/Time Discretization}	
In this subsection, we compare the trajectory representation of the proposed FPD scheme against the existing TD and CPD schemes. 

First, the following two lemmas compare the number of waypoints required for trajectory representation between TD and CPD, as well as between CPD and FPD, respectively.
\begin{lemma} \label{TimePath}\emph{
	Given $\{T, \Delta_{\rm max}, V_{\rm max}\}$, if the trajectory representation with TD (i.e., $\{\{{\bf q}_m\}, \delta_t\}$) satisfies:  1) the flying velocities over $A_1\ge 2$ consecutive line segments are the same, i.e., $\mathbf{v}_m=\mathbf{v}_{m+1}\cdots=\mathbf{v}_{m+A_1-1}={\bf v}_0$ for $1\le m\le M-(A_1+1)$; and 2) $\|\mathbf{v}_0\| \leq \frac{V_{\rm max}}{A_1}$, then CPD  requires $A_1-1$ fewer waypoints than TD, i.e., $(M+1) - (N+1) = A_1-1$. 
}
\end{lemma}
\begin{proof}
	Given the  conditions 1) and 2), the $A_1$ consecutive segments form a straight-line segment with a constant UAV speed $\|\mathbf{v}_{0}\|$ and a total segment length satisfying $\|\mathbf{q}_{m+A_1-1}-\mathbf{q}_{m-1}\|  \leq \|\mathbf{v}_{0}\| A_1 \delta_t  \leq \Delta_{\rm max}$. Thus,  these $A_1$ segments can be regarded as one segment for CPD, represented by the two waypoints $\{\mathbf{q}_{m-1},\mathbf{q}_{m+A_1-1}\}$, together with a UAV traveling duration $A_1\delta_t$.
	Hence, $K+1-2=K-1$ waypoints are saved for trajectory representation.
\end{proof}
\begin{lemma}\label{FlexVSPath}\emph{Given $\{T, \Delta_{\rm max}, V_{\rm max}\}$,  if the trajectory representation with CPD (i.e., $\{\{{\bf q}_n\}, \{t_n\}\}$) satisfies that the flying velocities over $A_2\ge 2$ consecutive line segments are the same, i.e., $\mathbf{v}_n=\mathbf{v}_{n+1}\cdots=\mathbf{v}_{n+A_2-1}$ for $1\le n\le N-(A_2+1)$, 
	 then the proposed FPD  requires $A_2-1$ fewer designable waypoints than CPD, i.e., $(N+1) - (L+1) = A_2-1$. }
\end{lemma}
Lemma~\ref{FlexVSPath} can be  proved by using the similar method as for proving Lemma~\ref{TimePath}, thus the details are omitted for brevity.
	
Lemma~\ref{TimePath} indicates that if the UAV hovers or moves slowly at the same direction and speed, CPD requires fewer (designable) waypoints than TD. 
Moreover, Lemma~\ref{FlexVSPath} shows that  compared to CPD, FPD in general can represent the UAV trajectory with even fewer (designable) waypoints as long as the UAV flies at the same velocity along (part of) its path, even at the maximum speed. For example, in the extreme case where the UAV flies  at a constant velocity towards the destination, FPD requires only $2$ designable waypoints (i.e., start and end locations) as well as its traveling duration for characterizing the trajectory, which in general is much smaller than that of CPD with $N\!+\!1\!=\!\l\lceil\frac{\|\mathbf{q}_{\rm str}\!-\!\mathbf{q}_{\rm end}\|}{\Delta_{\max}}\r\rceil+1$ waypoints. For ease of illustration, a concrete example is provided below for comparing the number of waypoints/time durations/design variables required  for trajectory representation under different trajectory discretization schemes.

	\begin{figure}[t] \centering    
	\includegraphics[width=0.90\columnwidth]{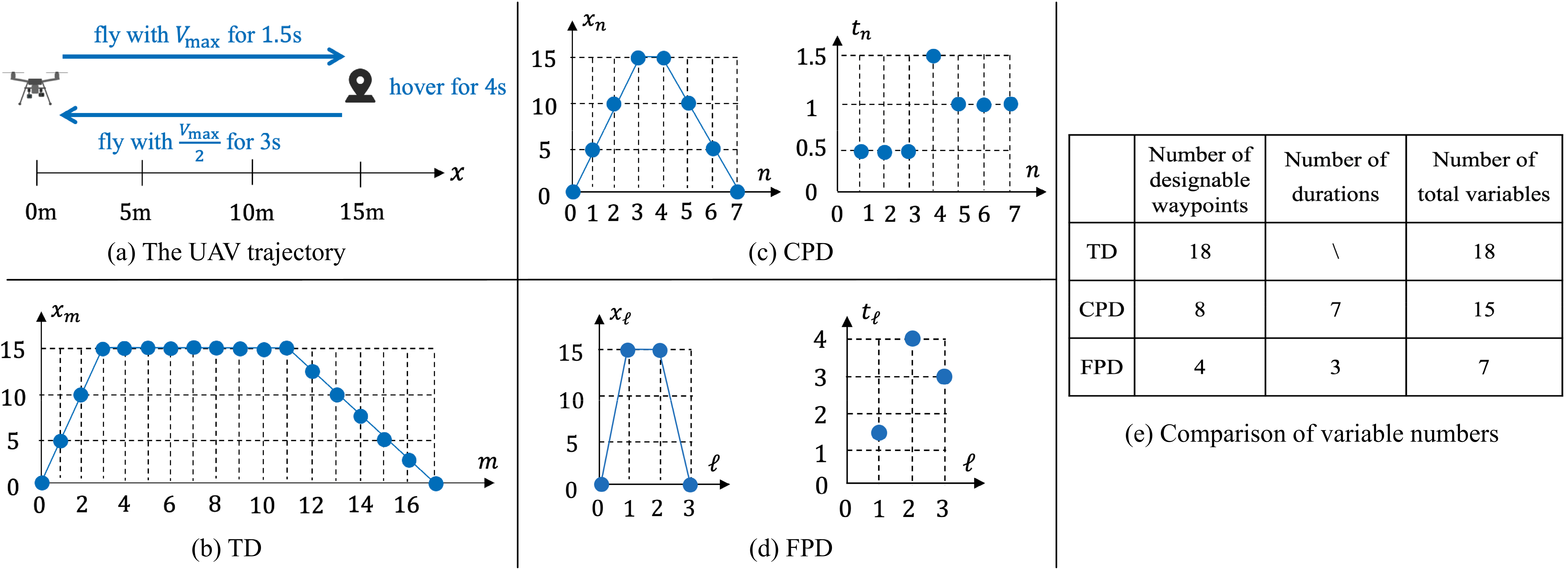} 
	\caption{{Trajectory representation under different trajectory discretization schemes.}}     
	\label{example}
	\end{figure}

\begin{example}\emph{
Consider an example illustrated in Fig.~\ref{example}(a), where the UAV flies along the $x$-axis. The 1D UAV trajectory consists of three consecutive  phases:  1) the UAV flies from $x=0$ m to $x=15$ m at the maximum speed of $V_{\rm max}=10$ m/s with a flight duration of $1.5$ s; 2) it  hovers above $x=15$ m  for $4$ s; 3) the UAV flies back to $x=0$ m at a constant speed of $\frac{V_{\rm max}}{2}=5$ m/s with a flight duration of  $3$ s.  We set $\Delta_{\rm max}=5$ m and have the following main observations. 
First, for TD, we have $\delta_t=\frac{\Delta_{\rm max}}{V_{\rm max}}=0.5$~s and the total flight period $T=8.5$ s. Hence,  TD requires $8.5/0.5+1=18$ waypoints in total for trajectory representation, as shown in Fig. \ref{example}(b).
	Second, for CPD, it requires only $2$ waypoints plus  the hovering time duration to represent  phase $2$ of UAV's hovering and $3$ waypoints for phase  $3$ with the UAV flying at a low speed; thus it involves $8$ waypoints and $7$ time durations in total as shown in Fig. \ref{example}(c).
	Last, for the proposed FPD with $J=3$, both phases $1$ and $3$ can be characterized by a long-segment with its corresponding flight duration since the UAV flies at a constant velocity. As such,  
 $4$ designable waypoints and $3$ time durations are sufficient for the trajectory representation, as shown in Fig.~\ref{example}(d).
In summary, the numbers of required design variables under different trajectory discretization schemes  are compared  in Fig.~\ref{example}(e).}
\end{example}	


\section{Proposed Path Compression}\label{SecCompre}
The FPD scheme proposed in the preceding section reduces the UAV trajectory design complexity in the \emph{time} domain via reducing the number of designable waypoints. In this section, we propose another alternative  approach, called \emph{path compression}, to further reduce the UAV trajectory design variables in  the \emph{spatial} domain.

\subsection{Path Decomposition}
Without loss of generality, we consider the UAV path  with FPD introduced in Section~\ref{SecFPD} as it includes TD and CPD as special cases, while the results can also  be extended to other trajectory discretization schemes. 
 Recall that with FPD, the UAV piecewise-linear trajectory can be characterized by the pre-determined parameters $\{L, J\}$, the  designable waypoints $\{{\bf q}_{\ell}\}$, as well as the corresponding time durations $\{t_{\ell}\}$. For this \emph{waypoint-based} path representation, the UAV path $\{{\bf q}_{\ell}\}$ is decomposed into $L+1$ waypoints and each waypoint ${\bf q}_{\ell}=[{q}_{\ell,{\rm x}},{q}_{\ell,{\rm y}},{q}_{\ell,{\rm z}}]^T$ is represented in the 3D Cartesian coordinate system.  Thereby, 
 the UAV trajectory is completely  represented by $3(L+1)$ waypoint variables  and $L$ time variables. In this subsection, we propose a new \emph{sub-path-based} path representation, for which the 3D UAV path  
 $\{{\bf q}_{\ell}\}$ is decomposed into three 1D sub-paths  in each of the 3D Cartesian coordinates, namely, $\{{\bf q}_{\ell}\}_{\ell=0}^{L}=\{{\bf q}_{\rm x},{\bf q}_{\rm y},{\bf q}_{\rm z}\}$ where ${\bf q}_{\rm dim}=[{q}_{0,{\rm dim}},\cdots,{q}_{L,{\rm dim}}]^T\in\mathbb{R}^{(L+1)\times 1}$ for $\dim={\rm x, y, z}$.
 Then, by  constructing $L+1$ \emph{basis paths} (or so-called \emph{path features}), we are able to represent each sub-path by a superposition  of the $L+1$ basis paths with flexibly chosen combining weights. For simplicity,  we consider in this paper that a  common set of basis paths are used  for each of the three sub-paths in 3D, while the results can be extended to the general case with different basis paths for different sub-paths.

	Define $\{{\bf p}_{\ell}\}_{\ell=0}^{L}$ as a set of  basis paths with ${\bf p}_{\ell}\in\mathbb{R}^{(L+1)\times 1}, \ell=0,\cdots, L$, provided that  ${\bf p}_{\ell}$' are linearly independent (as will be exemplified  later).
	With $\{{\bf p}_{\ell}\}_{\ell=0}^{L}$, the three sub-paths can be equivalently represented by
	\begin{align}
	({\bf q}_{\rm dim})^T=\sum_{\ell=0}^L c_{\ell,{\dim}} ~{\bf p}_{\ell}^T = ({\bf c}_{{\dim}})^T{\bf P}, \quad {\rm dim}={\rm x, y, z},\label{subpath}
	\end{align}
	where ${\bf c}_{{\dim}}=[c_{0,{\dim}}, \cdots, c_{L,{\dim}}]^T\in\mathbb{R}^{(L+1)\times 1}$ with $c_{\ell,{\dim}}$ denoting the path coefficient for basis path $\ell$ in the specified dimension, and ${\bf P}\triangleq[{\bf p}_{0}, \cdots, {\bf p}_{L}]^T\in\mathbb{R}^{(L+1)\times (L+1)}$ is named as the \emph{basis-path matrix}. Let $\mathbf{Q}=[\mathbf{q}_0,\cdots,\mathbf{q}_L]=[\mathbf{q}_{\rm x},\mathbf{q}_{\rm y},\mathbf{q}_{\rm z}]^T\in\mathbb{R}^{3\times (L+1)}$ denote the \emph{waypoint-matrix} that stacks all the $L+1$ designable waypoints. Based on the above, $\mathbf{Q}$ can be  expressed in the following compact form
	\begin{align}
	\mathbf{Q}= \mathbf{C}{\bf P},
	\end{align}
where $\mathbf{C}=[{\bf c}_{{\rm x}}, {\bf c}_{{\rm y}}, {\bf c}_{{\rm z}}]^T\in\mathbb{R}^{3\times (L+1)}$ denotes the \emph{path-coefficient} matrix. 
Since the basis paths are linearly independent,  $\mathbf{P}$ is of  full-rank and thus we have $\mathbf{C}=\mathbf{Q}\mathbf{P}^{-1}$. This means that under the new path coordinate system $\mathbf{P}$, the 3D UAV path $\mathbf{Q}$ can be fully  characterized by the path-coefficient matrix $\mathbf{C}$ with $3(L+1)$ variables that flexibly weight  the basis paths for constructing a desired path.

\begin{figure}[t] \centering    	
       \subfigure[Fourier basis paths.]
	{\label{PCFitExampleBasicA}     
	\includegraphics[width=0.45\columnwidth]{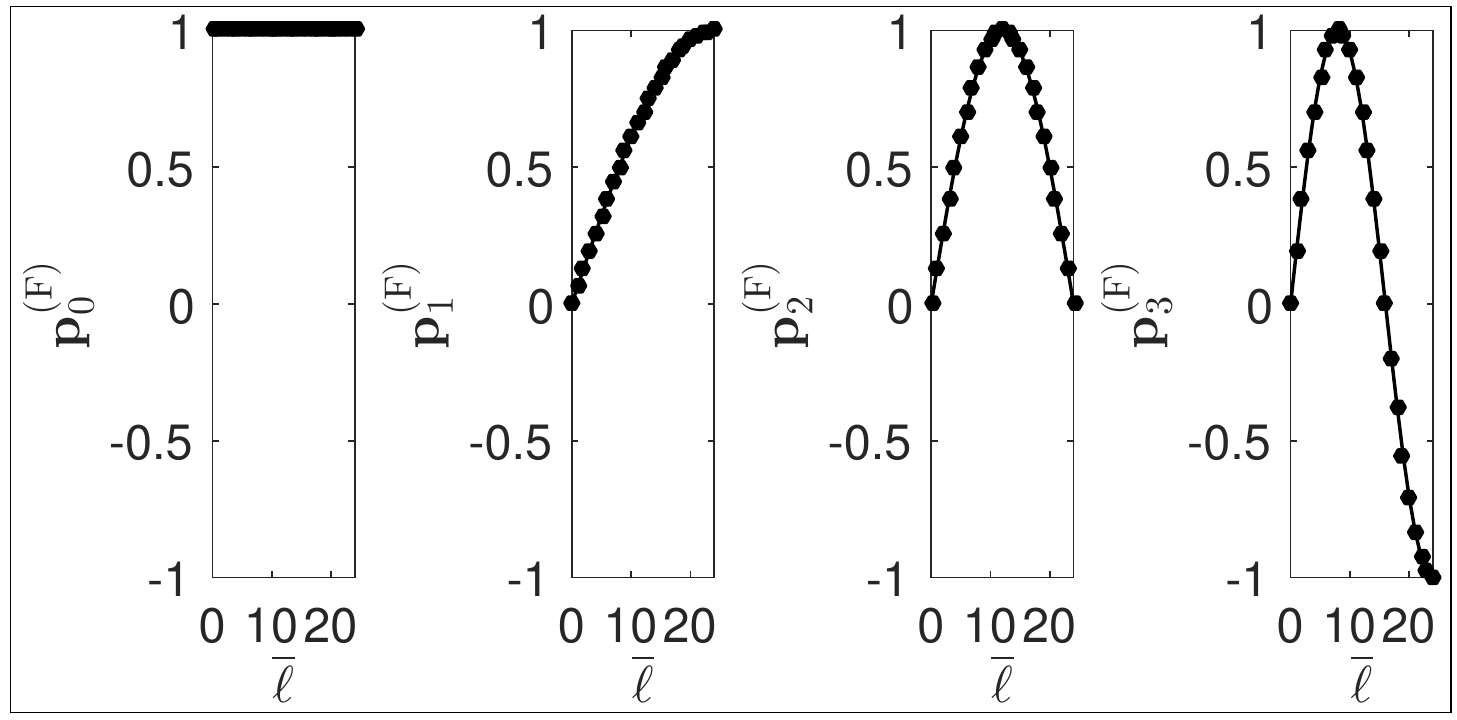} 
	}    
	\subfigure[Shifted-sine basis paths.]
	{\label{PCFitExampleBasicB}     
	\includegraphics[width=0.45\columnwidth]{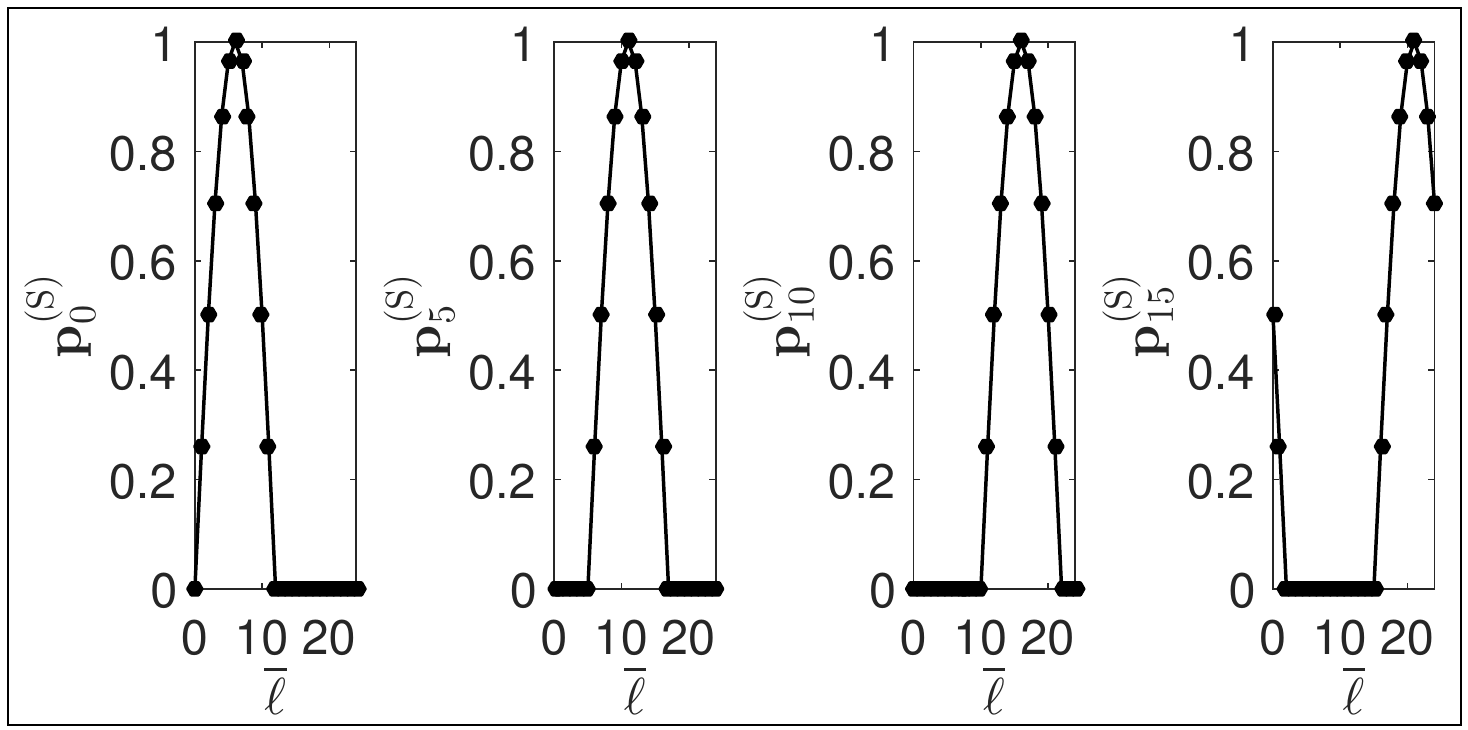} 
	} 
	\caption{Two example types of basis paths.}   
	\label{PCFitExampleBasic}
	\vspace{-10pt}
	\end{figure}
  
In general, there are various types of basis paths for ${\bf P}$, provided that they satisfy  $\rank({\bf P})=L+1$. In the following, we propose two example types of basis paths. 
\begin{example}[Fourier basis paths]\emph{Inspired by the Fourier decomposition for the continuous function, we define the Fourier basis paths  as ${\bf p}^{(\rm F)}_{\ell}=[{p}^{(\rm F)}_{\ell,0}, \cdots, {p}^{(\rm F)}_{\ell,L}]$ for $\ell=0, \cdots, L$, where 
\begin{align}
{p}^{(\rm F)}_{\ell,\tilde{\ell}}=
\begin{cases}
    	\sin\l(\frac{\pi \ell \tilde{\ell}}{2L}\r), & \tilde{\ell}=0, \cdots, L, \quad \ell=1, \cdots, L,\\
	1, & \rm otherwise,
\end{cases}
\label{Fourbasis}
\end{align}
Fig.~\ref{PCFitExampleBasicA} depicts some of the above defined Fourier basis paths for the case with $L=24$. It is observed that the basis-path variation over $\ell$ (or \emph{path frequency}) increases with the sub-path index $\ell$. Thus, one can draw an analogy between the path frequency and Fourier series, where the higher-order basis path corresponds to ``higher-frequency" component in each sub-path.}
\end{example}
\begin{example}[Shifted-sine basis paths]\label{Shifted}\emph{The shifted-sine basis paths are defined as  ${\bf p}^{(\rm S)}_{\ell}=[{p}^{(\rm S)}_{\ell,0}, \cdots, {p}^{(\rm S)}_{\ell,L}]$ for $\ell=0, \cdots, L$, where 
\begin{align}
{p}^{(\rm S)}_{\ell,\tilde{\ell}}=
\begin{cases}
	\sin\l(\frac{2\pi (\tilde{\ell}-\ell)}{L}\r), & \tilde{\ell}\in[0,b_1(\ell)]\cup[\ell,b_2(\ell)], \quad \ell=0, \cdots, L,\\
    	0, & \rm otherwise,\\
\end{cases}
\end{align}
with $b_1(\ell)=\max\{0, \ell-\frac{L}{2}-1\}$ and $b_2(\ell)=\min\{\frac{L}{2}+\ell, L\}$.
Fig.~\ref{PCFitExampleBasicB} plots some of the defined shifted-sine basis paths for the case with $L=24$. One can observe that  the  basis paths for different $\ell\ge1$ are shifted versions of ${\bf p}^{(\rm S)}_{0}$ and thus they have similar shapes.}
\end{example}

\vspace{-15pt}
\subsection{Path Compression} \label{SecPathComp}
Based on the Fourier basis paths proposed in the previous subsection as an example, we  propose in this subsection  a simple yet efficient method, named PC, to compress the UAV path in the spatial domain  and then reformulate the UAV trajectory optimization problem based on PC. Note that the proposed PC method can be similarly applied to other basis paths (e.g., shifted-sine basis paths in Example~\ref{Shifted}){\color{blue}.}  

The  main idea of PC is to approximate the UAV path as a  superposition  of  some selected Fourier basis paths only but still retain the main desired features of the path, thus achieving lower path design complexity yet with tolerable path-compression error. To be specific, let $K$ with $0<K\le L+1$ denote the  number of selected Fourier basis paths among the full $L+1$ basis paths (which fully characterize  the $L+1$ waypoints).  
We define $\rho_{\rm comp}=\frac{K}{L+1}\in(0,1]$ as the path-compression ratio. 
Since the UAV path is usually preferred to be  smooth in practice, we select the first $K$  basis paths that have relatively low path frequencies to construct a reduced-dimensional basis-path matrix ${\bar{\bf P}}\triangleq[\bar{\bf p}_{0}, \cdots, \bar{\bf p}_{K-1}]^T\in\mathbb{R}^{K\times (L+1)}$. Then,  the UAV path can be approximately characterized by the following compressed path-coefficient matrix $\bar{\mathbf{C}}$ 
\vspace{-5pt}
\begin{align}
	\mathbf{Q}\approx \bar{\mathbf{Q}}\triangleq \bar{\mathbf{C}}{\bar{\bf P}},
\end{align}
where $\bar{\mathbf{C}}=[\bar{\bf c}_{{\rm x}}, \bar{\bf c}_{{\rm y}}, \bar{\bf c}_{{\rm z}}]^T\in\mathbb{R}^{3\times K}$ with $\bar{\bf c}_{{\rm dim}}=[c_{0,{\dim}}, \cdots, c_{K-1,{\dim}}]^T\in\mathbb{R}^{K\times 1}, \dim={\rm x,y,z}$ denoting the compressed path coefficient vector for the sub-path in each corresponding dimension. Note that there exists a fundamental trade-off between the PC ratio (or path design complexity) and compression accuracy (or communication performance). Intuitively, as $K$ decreases, although fewer basis paths are selected which leads to a smaller number of path coefficients to be optimized, it also results in  less DoF in optimizing the UAV path that may cause communication performance degradation.

\begin{figure}[t] \centering    	
       \subfigure[PC based on Fourier basis paths.]
	{\label{PCFitExampleA}     
	\includegraphics[width=0.45\columnwidth]{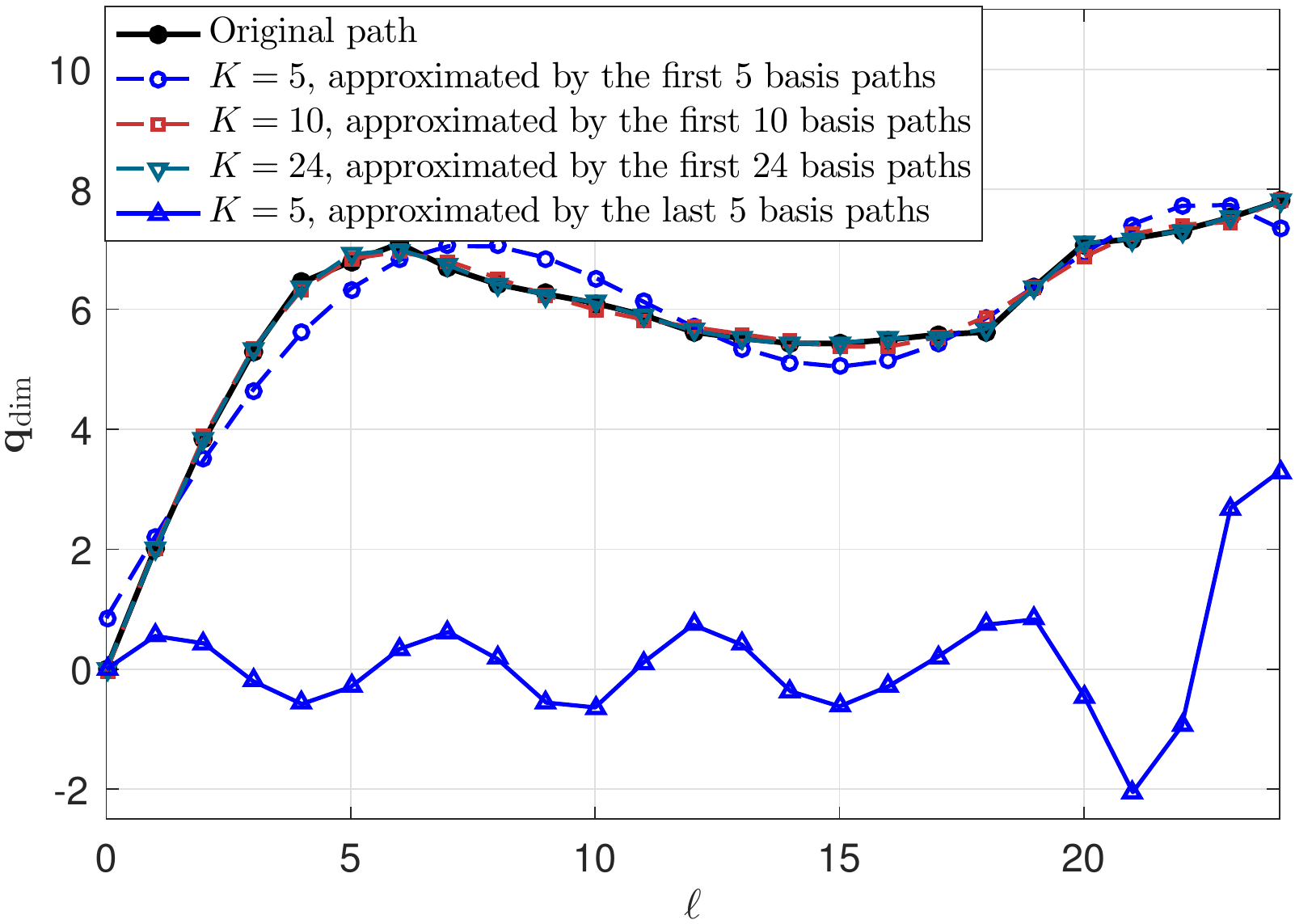} 
	}    
	\subfigure[PC based on shifted-sine basis paths.]
	{\label{PCFitExampleB}     
	\includegraphics[width=0.45\columnwidth]{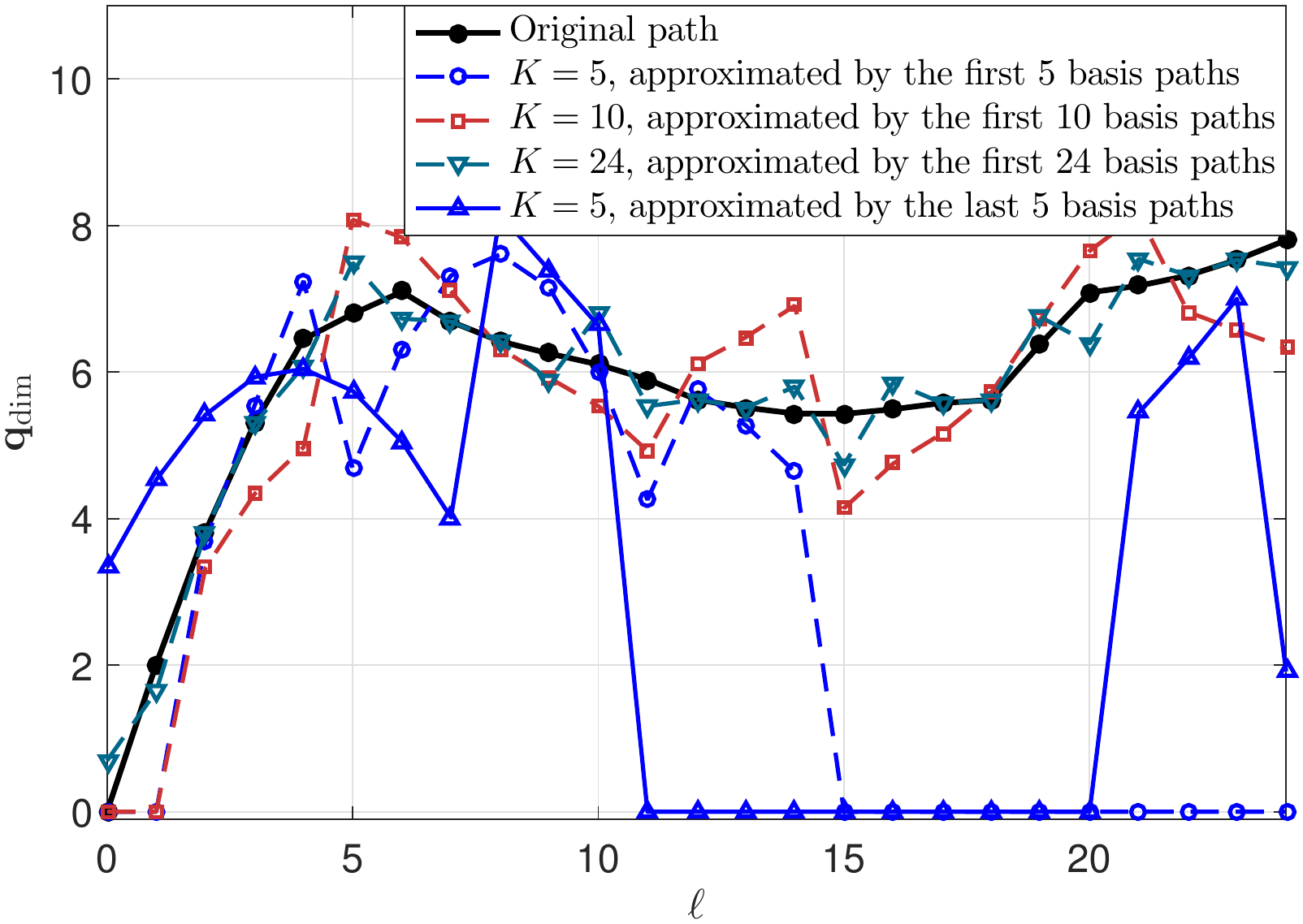} 
	} 
	\caption{Comparison of path approximation  for PC based on different basis paths.}   
	\label{PCFitExampleResult}
	\end{figure}

In Fig.~\ref{PCFitExampleResult}, we numerically compare the compressed path by the proposed Fourier-based PC with different $K$ against the benchmark scheme based on the shifted-sine basis paths introduced in Example~\ref{Shifted}. First, it is observed that for the proposed Fourier-based PC, as $K$ increases, the approximated UAV path gets closer to the actual path with more path basis components selected. Second, it is interesting to observe that for the path with $L+1=25$ waypoints, the proposed Fourier-based PC with small $K$ (e.g., $K=5$) achieves very close path approximation performance as that with large $K$ (e.g., $K=24$). This can be explained by the fact that the basis paths with low-order path frequencies contain most of information for this smooth  path in practice. Thus, discarding high-frequency basis paths does not compromise the approximation accuracy notably. Third, the proposed Fourier-based PC significantly outperforms the benchmark scheme based on the shifted-sine basis paths for different $K$, since the basis paths of the latter scheme are equally important  and thus the approximation accuracy is very sensitive to the number of (shifted) basis paths selected.

With the proposed Fourier-based PC, the designable waypoints with FPD can be rewritten as
${\bf q}_{\ell}=\bar{\bf C}[{\bar{\bf P}}]_{:,\ell+1}$, where $[{\bar{\bf P}}]_{:,\ell+1}$ denotes the $(\ell+1)$-th column of ${\bar{\bf P}}$. Thus, the communication utility function for the combined FPD with PC (termed FPD-PC) is given by 
\begin{align}
\bar{U}_{{\rm FPD-PC}}(\bar{\bf C},\{t_{\ell}\})=\bar{U}_{{\rm FPD}}(\{\bar{\bf C} [{\bar{\bf P}}]_{:,\ell+1}\},\{t_{\ell}\}),
\end{align}
where the function $\bar{U}_{{\rm FPD}}(\cdot)$ is defined in \eqref{Eq:FleSumAppro}. As such, problem ($\rm P4$) can be approximated by
\begin{subequations}
\begin{align} 
	\!\!\!\!\!{{\rm (P8)}:} ~\mathop{\max}_{\bar{\bf C},\{t_{\ell}\}}   \quad & \bar{U}_{{\rm FPD-PC}}(\bar{\bf C},\{t_{\ell}\})\nn \\
	{\rm s.t.} \quad
	&\bar{f}_{i, \rm{FPD-PC}}(\bar{\bf C},\{t_{\ell}\})\le 0, ~ i=1,\cdots,I,\label{Eq:FPD-PC}
\end{align} 
\end{subequations}
where $\{\bar{f}_{i,{\rm FPD-PC}}(\bar{\bf C},\{t_{\ell}\})\}$ denote the set of approximated constraints.
Note that compared to the FPD scheme without PC in Section~\ref{SecFPD} that requires $3(L+1) + L$ design variables, the new FPD-PC scheme has reduced the number of path design variables to $3K + L$ with $K\le L+1$.

\section{Case Study: UAV Trajectory Design for Min-rate Maximization}\label{Case}

A case study is provided in this section to show different formulations of the same min-rate maximization problem under different trajectory discretization schemes with/without  PC. Their computational complexities for solving their respective optimization problems are analyzed as well.

	Similar to Example~\ref{Example1}, we consider a UAV-enabled data harvesting system where a  UAV is dispatched to collect data from $S$ ground SNs, denoted by the set $\mathcal{S}=\{1,\cdots,S\}$. Let $\mathbf{w}_s\in\mathbb{R}^{3\times 1}, s\in \mathcal{S}$ denote the location of SN $s$ and we assume that the UAV starts its flight from $\mathbf{q}_{\rm str}$ and flies back to  $\mathbf{q}_{\rm end}=\mathbf{q}_{\rm str}$ after a flight period of $T$. For the purpose of exposition, we consider the LoS UAV-ground channel model \cite{tutor}, while the results can be extended to the more general UAV-ground channel models (e.g., the elevation-angle dependent Rician fading channel \cite{you20193d} and probabilistic LoS channel \cite{you2020hybrid}). Under this channel model, the UAV is assumed to fly at a fixed minimum UAV altitude $H_{\min}$. Moreover,  we consider time division multiple access (TDMA) for the UAV communication scheduling and denote by $\alpha_s(t)\in\{0,1\}$ the association variable for user $s$ at time $t$, where $\alpha_s(t)=1$ indicates that SN $s$ is scheduled by the UAV for transmission  and otherwise it keeps silent. As such, the average  data collection rate of the UAV from SN $s$ is given by
\begin{align}\label{R_k_continuous}
	R_s=\frac{1}{T}\int_0^T \alpha_s(t) \log_2\left( 1+ \frac{P_s\beta_0}{\|\mathbf{q}(t)-\mathbf{w}_s\|^2\sigma^2} \right) dt,
	\end{align}
where $P_s$ denotes the transmit power of SN $s$, and  $\{\beta_0,\sigma^2\}$ are defined in Example~\ref{Example1}.
	
	 Our objective is to maximize the minimum (average) achievable rate among all the $S$ ground SNs by jointly designing the UAV continuous-time piecewise-linear trajectory and communication scheduling, i.e., $U(\mathcal{Q}(t),\mathcal{R}(t))=\min_{s\in\mathcal{S}}R_s$, where $\mathcal{Q}(t)=\mathbf{q}(t)$ and $\mathcal{R}(t)=\{\alpha_s(t)\}_{s=1}^S$ in $(\rm P1)$.
	Under the practical constraints on the UAV trajectory and communication scheduling, the above min-rate maximization problem is  formulated as
	\begin{subequations}
\begin{align} 
	\!\!\!\!\!{{\rm (P9)}:} ~\mathop{\max}_{\mathbf{q}(t), \{\alpha_s(t)\}}   \quad & \min_{s\in\mathcal{S}}R_s\nn \\
	{\rm s.t.} \qquad
	&\label{consSpeed_I}\|\dot{\mathbf{q}}(t)\| \leq V_{\max},~ {q}_{\rm z}(t)=H_{\min}, ~\forall  t\in[0,T],\\
	&\label{consStart_I} \mathbf{q}(0)=\mathbf{q}_{\rm str},~ \mathbf{q}(T)=\mathbf{q}_{\rm end},\\
	&\label{consConnect_I}\sum_{s=1}^S \alpha_s(t) \leq 1, ~\forall t\in[0,T],\\
	&\label{cons01_I}\alpha_s(t) \in \{0,1\}, ~ \forall s\in\mathcal{S}, ~t\in[0,T],
\end{align} 
\end{subequations}
where (\ref{consSpeed_I}) imposes the constraints on the UAV  maximum  speed and fixed altitude, (\ref{consStart_I}) specifies the UAV start/end locations, 
 (\ref{consConnect_I}) restricts the UAV scheduling with one user only at each time instant,  and (\ref{cons01_I}) denotes the binary communication scheduling constraint for $\alpha_s(t)$ over $t$. 

\subsection{Problem Reformulation}
\subsubsection{TD}
	For the TD scheme, we set $\delta_t=\Delta_{\max}/V_{\max}$  given a properly chosen $\Delta_{\max}$ and thus $M=T/ \delta_t$ (assumed to be an integer). Similar to \cite{b5}, the binary communication scheduling constraint for each time slot $m$ in \eqref{cons01_I} is relaxed as  $0\le\alpha_{s,m}\le1, \forall s\in\mathcal{S}, m=1, \cdots, M$. Then the approximated communication utility under the TD scheme is given by $\bar{U}_{\rm TD}(\{{\bf q}_m\}, \{\alpha_{s,m}\})=\min_{s\in\mathcal{S}}\bar{R}_{s,{\rm TD}}$, where $\bar{R}_{s,{\rm TD}}=\frac{1}{M}\sum_{m=1}^{M} \alpha_{s,m} \log_2 \left(1+\frac{P_s\beta_0}{\|\mathbf{q}_m-\mathbf{w}_s\|^2\sigma^2}\right).$
	As such, with TD,  problem ($\rm P9$) is reformulated as
\begin{subequations}
\begin{align} 
	{\rm{(P10)}:}\quad \mathop{\max}_{\{\mathbf{q}_m\}, \{\alpha_{s,m}\}}   \quad & \min_{s\in\mathcal{S}}\bar{R}_{s,{\rm TD}} \nn\\
{\rm s.t.}~ \qquad
	&\label{consSpeed_TD}\|\mathbf{q}_m-\mathbf{q}_{m-1}\| \leq \Delta_{\max},  ~ m=1,\cdots, M,\\
	&\label{consStart_TD} \mathbf{q}_0=\mathbf{q}_{\rm str},~\mathbf{q}_M=\mathbf{q}_{\rm end},~q_{m,{\rm z}}=H_{\min},  ~ m=0,\cdots, M,\\
	&\label{consConnect_TD}\sum_{s=1}^S \alpha_{s,m} \leq 1, ~0\le \alpha_{s,m} \le1, ~\forall s\in\mathcal{S},   ~m=1, \cdots, M.
\end{align} 	
\end{subequations}
Note that given fixed UAV altitude, we only need to optimize the 2D horizontal  trajectory for $\{\mathbf{q}_m\}$.

\subsubsection{CPD}

With CPD and given the same $\Delta_{\max}$ as for TD, the UAV trajectory can be characterized by  $\{\{\mathbf{q}_n\}_{n=0}^{N},\{t_n\}_{n=1}^{N}\}$. Following the procedures in Section~\ref{SecPath}, the communication utility function can be approximated by
$\bar{U}_{\rm CPD}(\{{\bf q}_n\},\{t_n\}, \{\alpha_{s,n}\})=\min_{s\in\mathcal{S}}\bar{R}_{s,{\rm CPD}}$, where $\bar{R}_{s,{\rm CPD}}=\frac{1}{T}\sum_{n=1}^{N} \alpha_{s,n} t_n \log_2 \left(1+\frac{P_s\beta_0}{\|\mathbf{q}_n-\mathbf{w}_s\|^2\sigma^2}\right)$.
As such, problem ($\rm P9$) can be approximately reformulated as 
\begin{subequations}
\begin{align} 
	\!\!\!\!\!\!{\rm{(P11)}:}~ \mathop{\max}_{\{\mathbf{q}_n\}, \{t_{n}\},  \{\alpha_{s,n}\}}   \quad & \min_{s\in\mathcal{S}}\bar{R}_{s,{\rm CPD}} \nn \\
{\rm s.t.}~~ \qquad &\label{consSpeed_CPD}\|\mathbf{q}_n-\mathbf{q}_{n-1}\| \leq \min\{\Delta_{\max}, t_nV_{\max}\},~ n=1,\cdots, N,\\
	&\label{consStart_CPD} \mathbf{q}_0=\mathbf{q}_{\rm str},~ \mathbf{q}_N=\mathbf{q}_{\rm end}, ~ q_{n,{\rm z}}=H_{\min},~ n=0,\cdots, N,\\
	&\label{consConnect_CPD}\sum_{s=1}^S \alpha_{s,n} \leq 1, ~0\le \alpha_{s,n}\le 1, ~\forall s\in\mathcal{S}, ~ n=1, \cdots, N,\\
	&\label{cons01_CPD} \sum_{n=1}^{N}t_n\leq T, ~n=1, \cdots, N.
\end{align} 	
\end{subequations}

\subsubsection{FPD}
Note that  for the proposed FPD scheme with parameter $\{L, J\}$, the time duration of the $j$-th short-segment of long-segment $\ell$ (i.e., $t_{\ell}/J$) is shared by all the users. Thus, the approximated communication utility is given by
$\bar{U}_{\rm FPD}(\{{\bf q}_{\ell}\},\{t_{\ell}\}, \{ \alpha_{s,\ell}[j] \})=\min_{s\in\mathcal{S}}\bar{R}_{s,{\rm FPD}}$, where 
$	\bar{R}_{s,{\rm FPD}}=\frac{1}{T}\sum_{\ell=1}^{L} \sum_{j=1}^{J} \alpha_{s,\ell}[j] \frac{t_\ell}{J} \log_2 \left(1+\frac{P_s\beta_0}{\|\mathbf{q}_{\ell}[j]-\mathbf{w}_s\|^2\sigma^2}\right)$
	with $\mathbf{q}_{\ell}[j]=\mathbf{q}_{\ell}+\frac{j(\mathbf{q}_{\ell}-\mathbf{q}_{\ell-1})}{J}$ and $\alpha_{s,\ell}[j]$ denoting the communication scheduling variable of user $s$ at the $j$-th short-segment of long-segment $\ell$.   As such, with FPD, problem ($\rm P9$) can be approximately formulated as	
\begin{subequations}
\begin{align} 
	\!\!\!{\rm{(P12)}:}~ \mathop{\max}_{\{{\bf q}_{\ell}\},\{t_{\ell}\}, \{ \alpha_{s,\ell}[j] \}}   \quad & \min_{s\in\mathcal{S}}\bar{R}_{s,{\rm FPD}} \nn \\
{\rm s.t.}~~~ \qquad	&\label{consSpeed_CPD}\|\mathbf{q}_{\ell}-\mathbf{q}_{\ell-1}\| \leq \min\{J\Delta_{\max}, t_{\ell}V_{\max}\},  ~ \ell=1, \cdots, L,\\
	&\label{consStart_CPD} \mathbf{q}_0=\mathbf{q}_{\rm str},~ \mathbf{q}_L=\mathbf{q}_{\rm end}, ~q_{\ell,{\rm z}}=H_{\min}, ~ \ell=0, \cdots, L,\\
	&\label{consConnect_CPD}\sum_{s=1}^S \alpha_{s,\ell}[j] \leq 1, ~0\leq\alpha_{s,\ell}[j] \leq 1, ~ \forall s\in\mathcal{S}, ~j=1, \cdots, J, ~\ell=1, \cdots, L,\\
	&\label{consTn_CPD}\sum_{\ell=1}^{L}t_{\ell}\leq T, ~\ell=1, \cdots, L.
\end{align} 	
\end{subequations}

\subsubsection{FPD-PC}
Recall that for PC, we construct the reduced basis-path matrix ${\bar{\bf P}}$ based on the first $K$ Fourier basis paths  in \eqref{Fourbasis} and the designable waypoints are functions of the reduced path coefficients, i.e.,  ${\bf q}_{\ell}=\bar{\bf C}[{\bar{\bf P}}]_{:,\ell+1}, \ell=0, \cdots, L$. Thus, the approximated communication utility  is given by $\bar{U}_{\rm FPD-PC}(\bar{\bf C},\{t_{\ell}\}, \{ \alpha_{s,\ell}[j] \})=\min_{s\in\mathcal{S}}\bar{R}_{s,{\rm FPD-PC}}$, where  $\bar{R}_{s,{\rm FPD-PC}}$ has a similar form as $\bar{R}_{s,{\rm FPD}}$ but with  ${\bf q}_{\ell}$ replaced by $\bar{\bf C}[{\bar{\bf P}}]_{:,\ell+1}$.
Note that given fixed UAV altitude $H_{\min}$, we only need to optimize the UAV horizontal trajectory and thus only the first two rows of the compressed coefficient matrix, i.e., $[\bar{\bf C}]_{1:2,:}$. 
As such, problem ($\rm P9$) is approximated by
%
\begin{subequations}
\begin{align} 
\!\!\!\!\!\!\!{\rm{(P13)}:}\mathop{\max}_{[\bar{\bf C}]_{1:2,:},\{t_{\ell}\}, \{ \alpha_{s,\ell}[j]\}} ~~ & \min_{s\in\mathcal{S}}\bar{R}_{s,{\rm FPD-PC}}\nn\\
{\rm s.t.}~~~ \qquad	
&\label{consSpeed_GPCD}\|[\bar{\bf C}]_{1:2,:}[{\bar{\bf P}}]_{:,\ell+1}-[\bar{\bf C}]_{1:2,:}[{\bar{\bf P}}]_{:,\ell}\| \leq \min\{J\Delta_{\max}, t_{\ell}V_{\max}\},    ~ \ell=1, \cdots, L,\\
	&\label{consStart_GPCD} [\bar{\bf C}]_{1:2,:}[{\bar{\bf P}}]_{:,1}=[\mathbf{q}_{\rm str}]_{1:2}, ~ [\bar{\bf C}]_{1:2,:}[{\bar{\bf P}}]_{:,L+1}=[\mathbf{q}_{\rm end}]_{1:2},\\
	\nonumber &\eqref{consConnect_CPD}-\eqref{consTn_CPD}.
\end{align} 	
\end{subequations}

Problems ($\rm P10$)--($\rm P13$) are non-convex optimization problems due to the coupling among the 2D UAV waypoints, traveling durations on line segments, as well as UAV communication scheduling. To tackle this difficulty, we can use the BCD method to decouple the variables into multiple blocks. For example, for the case with FPD, problem ($\rm P12$) can be decomposed into three subproblems, corresponding to the optimization for the blocks of $\{{\bf q}_{\ell}\},\{t_{\ell}\}$, and $\{ \alpha_{s,\ell}[j]\}$, respectively. Although the subproblems associated with $\{t_{\ell}\}$ and $\{ \alpha_{s,\ell}[j]\}$ can be optimally solved owing to their convexity, the optimization problem for $\{{\bf q}_{\ell}\}$ in general is still non-convex and thus difficult to solve. Fortunately, it has been shown in the existing literature (see, e.g., \cite{tutor}) that the SCA technique can be utilized to transform this kind of non-convex optimization problem into solving a series of relaxed convex optimization problems so as to obtain its suboptimal solution efficiently. The details are  thus omitted for brevity.

\subsection{Proposed Algorithms and Complexity Analysis}
\begin{table}[t]\small
\centering
\begin{spacing}{1.15}
\begin{tabular}{|c|c|c|c|c|}
\hline
\multirow{2}{*}{Optimization problem} & \multicolumn{2}{c|}{Trajectory design} &\multicolumn{2}{c|}{Communication design}\\\cline{2-5}
 & $N_{\rm{trj}}$ & Complexity order& $N_{\rm{com}}$ & Complexity order\\\hline
($\rm P10$): TD	   	       & $2(M+1)$ & $\mathcal{O}(M^{3.5}\log(1/\epsilon))$ & $UM$ &  $\mathcal{O}((UM)^{3.5}\log(1/\epsilon))$ \\\hline
($\rm P11$): CPD		       & $2(N+1)$ & $\mathcal{O}(N^{3.5}\log(1/\epsilon))$ & $UN+N$ &  $\mathcal{O}((UN+N)^{3.5}\log(1/\epsilon))$ \\\hline
($\rm P12$): FPD	& $2(L+1)$ & $\mathcal{O}(L^{3.5}\log(1/\epsilon))$ & $ULJ+L$ &  $\mathcal{O}((ULJ+L)^{3.5}\log(1/\epsilon))$ \\\hline
($\rm P13$): FPD-PC	&  $2K$ & $\mathcal{O}(K^{3.5}\log(1/\epsilon))$ & $ULJ+L$ &  $\mathcal{O}((ULJ+L)^{3.5}\log(1/\epsilon))$ \\\hline
\end{tabular}
\end{spacing}
\caption{Comparison of computational complexities of different trajectory discretization schemes with/without PC. Note that $K\leq L\leq N\leq M$ in general.}
\label{CompleComp}
\end{table}

Instead, we focus on analyzing  the computational complexity of the algorithms based on the BCD and SCA with different UAV trajectory discretization schemes. Specifically, the 2D trajectory (or waypoints) optimization can be solved by the standard interior-point method with the complexity order of $\mathcal{O}(N_{\rm{trj}}^{3.5}\log(1/\epsilon))$ \cite{ben2001lectures}, 
where $N_{\rm{trj}}$ denotes the number of scalar variables for trajectory optimization and $\epsilon>0$ is the solution accuracy.
On the other hand, the communication design problem is an LP in our considered problem  and has the (worst-case) complexity order of 
$\mathcal{O}(N_{\rm{com}}^{3.5}\log(1/\epsilon))$ \cite{ben2001lectures},
where $N_{\rm{com}}$ denotes the number of variables relevant to the communication design. Notice that although  the trajectory design has the same complexity scaling order as the communication counterpart in terms of number of scaler variables, it involves solving a second-order cone programming (SOCP) which usually takes much longer running time than solving an LP for the communication design by using well-known optimization softwares e.g., CVX \cite{grant2008cvx};
 as such, their total complexity is generally dominated by the trajectory design, when the number of variables for both designs are practically large.   
Based on the above, we summarize in Table~\ref{CompleComp} the computational complexity required  for solving  problems ($\rm P10$)--($\rm P13$) with TD, CPD, FPD, and FPD-PC, respectively. It is observed that the UAV trajectory and communication co-design with the proposed FPD scheme has lower total complexity as compared to TD and CPD, while the proposed FPD-PC scheme further reduces the complexity of FPD (without PC). The communication performance of these UAV trajectory designs will be evaluated in the next section by simulation.

\vspace{-5pt}
\section{Numerical Results}\label{Simulation}

 Numerical results are presented in this section to verify the effectiveness of the proposed FPD and PC schemes. We consider a UAV-enabled data harvesting system with $10$ ground SNs randomly and uniformly distributed in a square area of $100\times100$ m$^2$. For ease of illustration, the following results are based on one specific realization of SNs' locations. The UAV starts its flight from the location $\mathbf{q}_{\rm str}=[0,0,100]^T$  and flies back to $\mathbf{q}_{\rm end}=\mathbf{q}_{\rm str}$ after a flight period of $T=100$ s, given a fixed (minimum) UAV flight altitude of $H_{\min}=100$ m and maximum UAV (horizontal) speed of $V_{\max}=20$~m/s. The tolerable finite-sum approximation error for the communication utility is set as $E_u=0.05$ bps/Hz. As such, we obtain $\Delta_{\max}\leq5.41$ m from Lemma 1 and hence set $\Delta_{\max}=5$ m for simplicity.
 For TD, the time-slot length is obtained as $\delta_t=\Delta_{\max}/V_{\max}=0.25$ s and the number of time slots is $M=T/\delta_t=400$.  Moreover, the channel power gain at the reference distance of $1$ m is set as $\beta_0=-60$ dB, the receiver noise power is $\sigma^2=-90$ dBW, and all the SNs' are assumed to send data at the same transmit power $P_s=0.2$ W, $\forall s\in\mathcal{S}$. All simulations are run in MATLAB 2016b, which operates on a computer equipped with Intel-i7, $3.5$ Hz processor, and $16$ GB RAM memory.

\subsection{Performance  of Proposed Flexible Path Discretization}


	\begin{figure}[t] \centering 
	\subfigure[Max-min rate versus number of (short-) segments.]  
	{\includegraphics[width=0.38\columnwidth]{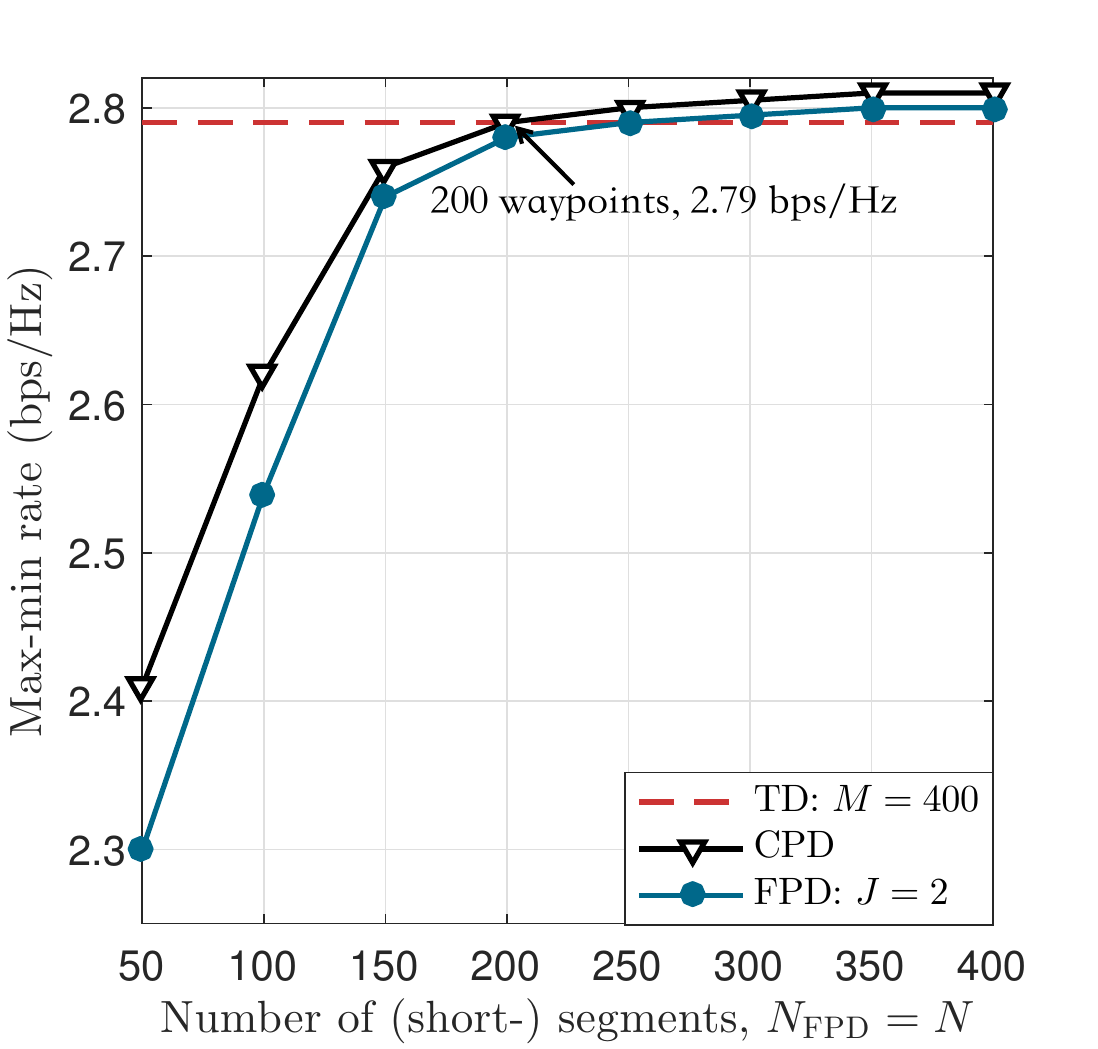} 
	\label{FigA_obj_N}
	}
\hspace{15mm}
	\subfigure[Running time versus number of (short-) segments.]  
	{\includegraphics[width=0.38\columnwidth]{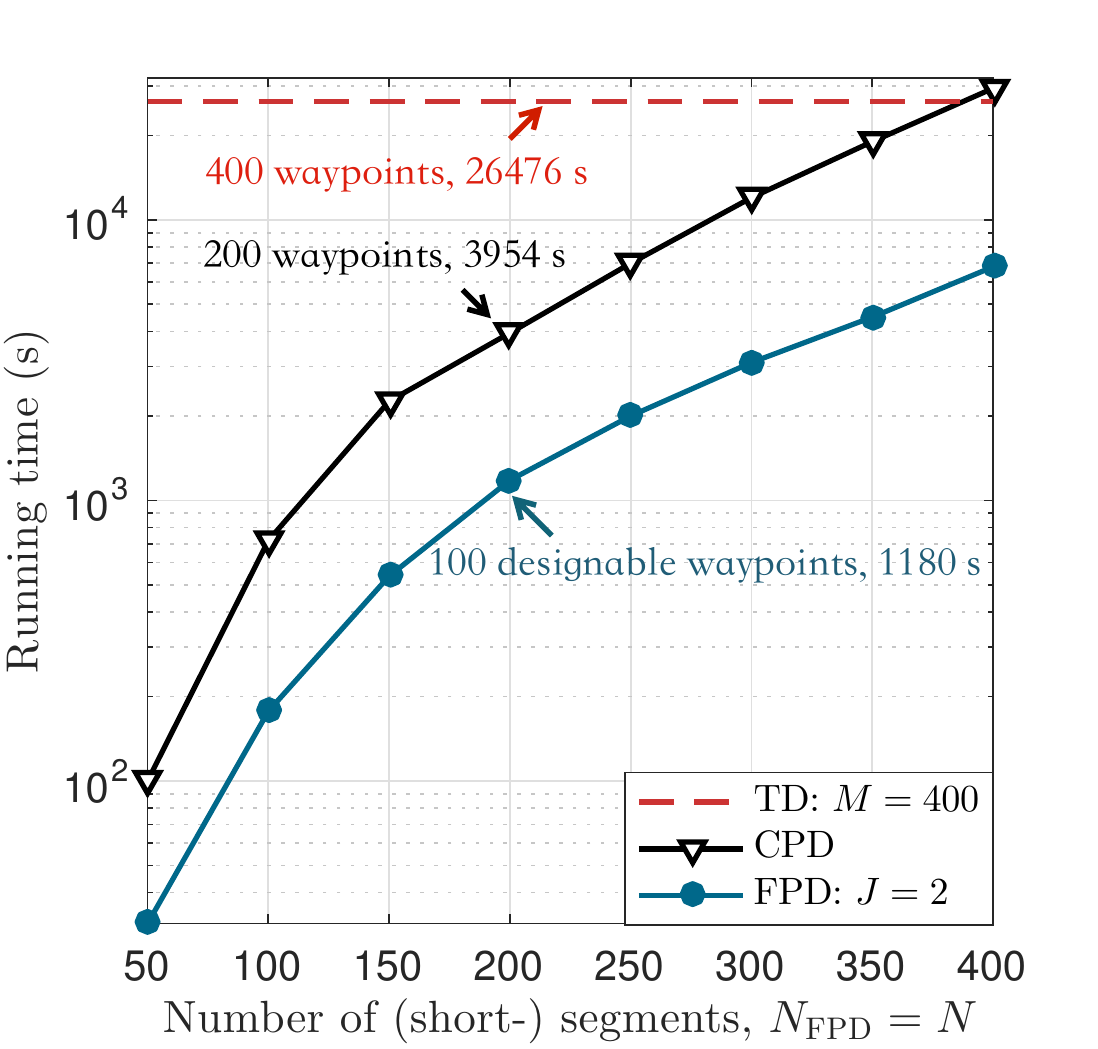}  
	\label{FigA_complex_N}
	}
	\caption{Performance comparison of different trajectory discretization schemes.}   
	\end{figure}

	In Figs.~\ref{FigA_obj_N} and \ref{FigA_complex_N}, we compare the achievable max-min rate and algorithm running time by the proposed FPD scheme with $J=2$ (i.e., $2$ short-segments in each long-segment) versus its number of short-segments, $N_{\rm FPD}$, 
	 against the TD and CPD benchmark schemes. Notice that for comparison, the horizontal axis represents the number of segments for the CPD benchmark ($N$) as well.
	  The main observations are given as follows.
First, it is observed  that as $N_{\rm FPD}$ increases, the max-min rate of the proposed FPD scheme firstly grows fast and then saturates after $N_{\rm FPD}$ exceeds a certain threshold  (see Fig.~\ref{FigA_obj_N}), while its running time monotonically increases with $N_{\rm FPD}$.  
The diminishing rate improvement with $N_{\rm FPD}$ can be explained by the fact that  increasing the number of designable waypoints significantly improves  the UAV trajectory design when $N_{\rm FPD}$ is small, while the improvement  becomes marginal when $N_{\rm FPD}$ is relatively large as it only results in small trajectory adjustment.
 The same rate and running time trends apply to the CPD scheme as well since it can be regarded as a special case of the FPD scheme with $J=1$.
Second, comparing the three trajectory discretization schemes, 
one can observe that the CPD scheme achieves smaller max-min rate than the TD scheme when $N$ is small, but slightly outperforms the latter when $N>200$ with even shorter algorithm running time.
 Moreover, compared to the CPD scheme, 
the proposed FPD scheme given the same number of  (short-) segments (i.e., $N_{\rm FPD}=N$) achieves very close max-min rate as the CPD scheme when $N_{\rm FPD}=N\ge 200$ (see Fig.~\ref{FigA_obj_N}), while it takes much shorter running time of $1180$ s versus $3954$ s of the CPD (see Fig.~\ref{FigA_complex_N}). However,  the proposed FPD scheme suffers more severe rate performance loss than the CPD scheme when $N_{\rm FPD}$ is small. On the other hand, given the same number of designable waypoints for both the FPD and CPD schemes (i.e., $N_{\rm FPD}=2N$), the former scheme significantly outperforms the latter in terms of achievable rate (see Fig.~\ref{FigA_obj_N}) while at the cost of comparable running time (see Fig.~\ref{FigA_complex_N}). This is because given $N_{\rm FPD}=2N$, the UAV trajectory by the FPD scheme is composed of  more waypoints (including both designable and non-designable ones) than the CPD scheme, which leads to longer path length in general and thus more DoF in the UAV trajectory design.

\begin{figure}
	\hspace{9mm}
	\subfigure[Max-min rate versus $J$.]
	{\includegraphics[width=0.42\columnwidth]{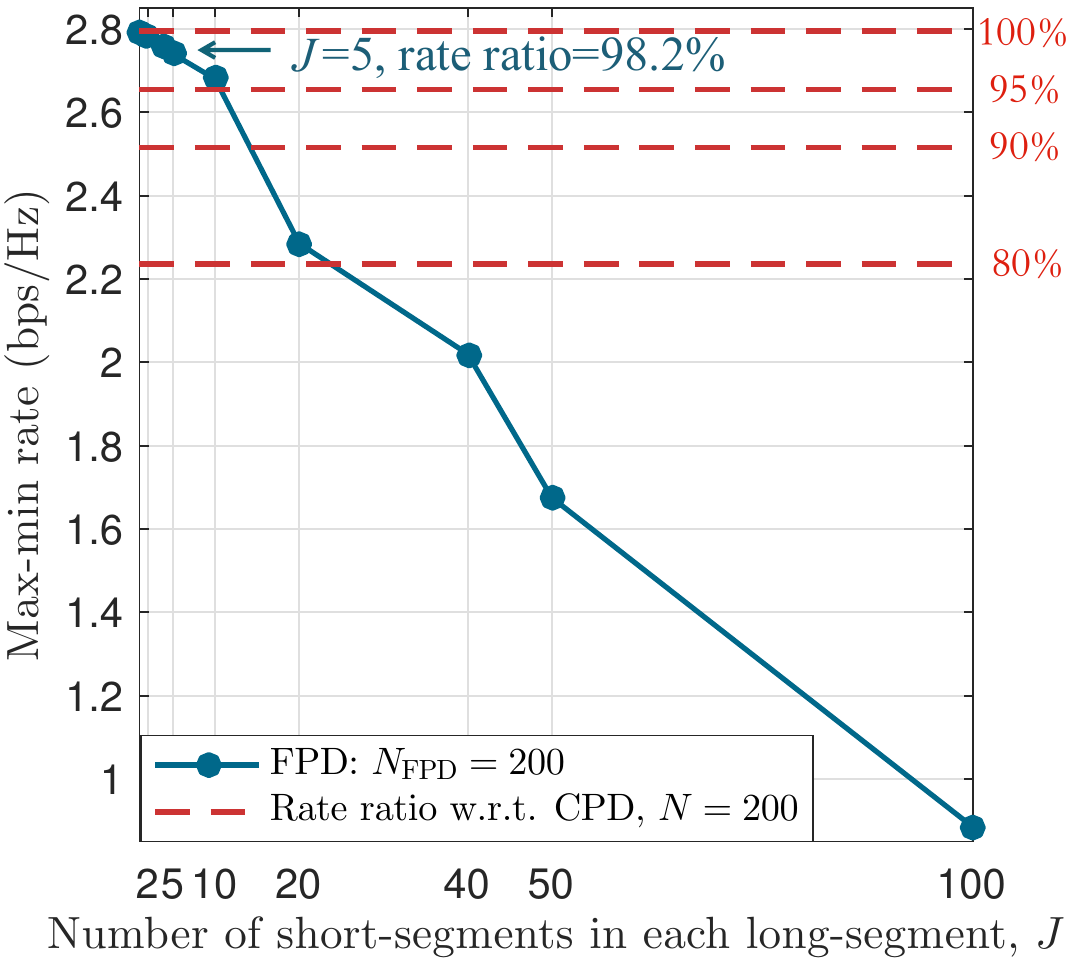} 
	\label{FigA_obj_J}
	}
	\hspace{15mm}
	\subfigure[Running time versus $J$.]
	{\includegraphics[width=0.42\columnwidth]{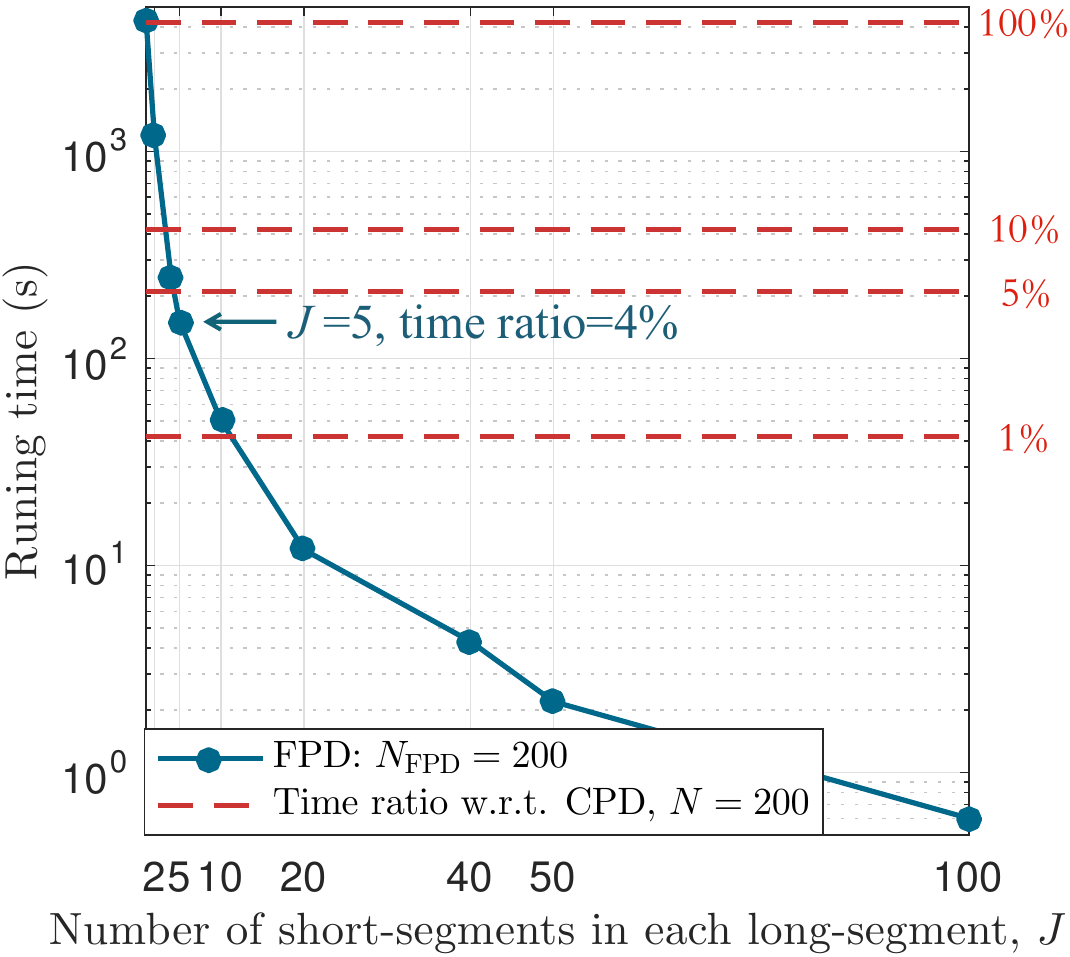} 
	\label{FigA_complex_J}
	}
	\caption{Effects of parameter $J$ on the performance of the proposed FPD scheme.}     
	\end{figure}

Figs.~\ref{FigA_obj_J} and \ref{FigA_complex_J} show the effects of parameter $J$ (i.e., number of short-segments in each long-segment) on the max-min rate and running time for the proposed FPD scheme given a fixed number of total short-segments $N_{\rm FPD}=200$. One can observe  that   as $J$ increases, the  max-min rate of the proposed FPD scheme monotonically decreases (see Fig.~\ref{FigA_obj_J}), while its algorithm running time firstly drops quickly with $J$ and then reduces slowly when $J$ becomes large. This is expected since with larger $J$, fewer designable waypoints need to be optimized which leads to both rate performance degradation and running time reduction due to less DoF in the UAV trajectory design.
Moreover, note that compared to the CPD scheme with the same number of total waypoints $N+1=201$, 
 the proposed FPD scheme with $J=5$ and hence only $41$ designable waypoints attains more than $98\%$ of the achievable rate of the CPD scheme, while it takes only $4\%$ of the  running time of the CPD scheme. Such a result  is appealing for practical implementation, which suggests that we can set small $J$ for the FPD scheme to reap most of the rate performance by the CPD scheme but  with substantially reduced running time.

	\begin{figure} \centering 
	\includegraphics[width=1\columnwidth]{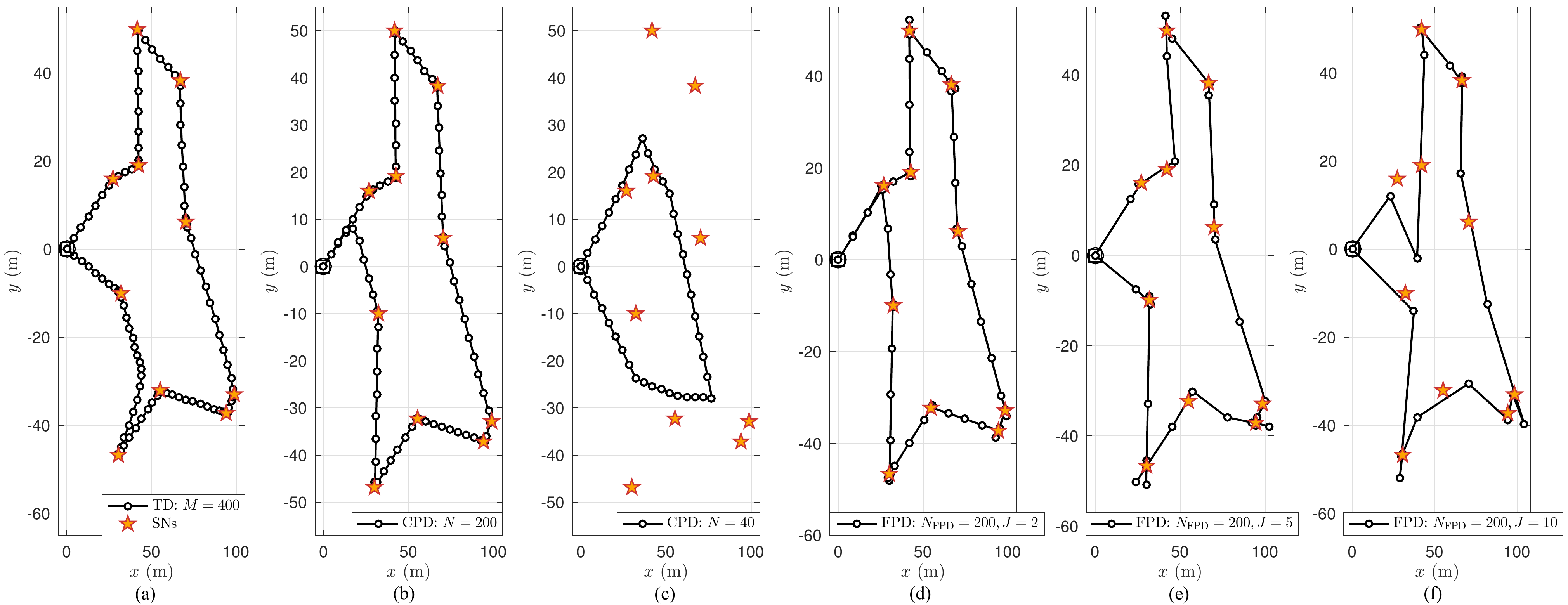} 
	\caption{Optimized UAV trajectories by different trajectory discretization schemes.}     
	\label{FigA_optTrjs}
	\end{figure}
	
	Fig.~\ref{FigA_optTrjs} plots the optimized UAV trajectories by different trajectory discretization schemes. First, it is observed that the optimized UAV trajectory by the CPD scheme with $N=200$ segments is similar to that of the TD scheme with much more segments, $M=400$, both of which  tend to sequentially travel nearby each SN (resembling the traveling-salesman-problem (TSP) solution \cite{tutor}). 
Second, as $J$ decreases, the optimized UAV trajectory by the proposed FPD scheme resembles that of the CPD scheme in the overall trajectory shape more closely.  Specifically, with smaller $J$, the UAV trajectory is characterized by more designable waypoints, which allows the UAV to more flexibly control its trajectory nearby the SNs. Third, comparing Figs.~\ref{FigA_optTrjs}(b) and \ref{FigA_optTrjs}(e), one can observe that given the same (small) number of designable waypoints (i.e., $N_{\rm FPD}/J+1=N+1=41$) and hence comparable algorithm running time, the optimized UAV trajectory by the proposed FPD scheme has more DoF to fly nearby the SNs for data collection than the CPD benchmark, since its trajectory consists of more (short-) segments and thus generally longer path length, which leads to larger max-min rate than the CPD scheme (cf. Fig.~\ref{FigA_obj_N}).

\subsection{Performance of Proposed  Path Compression}

	\begin{figure} \centering 
	\subfigure[Max-min rate versus number of selected basis paths.] 
	{\label{FigB_obj_K}
	\includegraphics[width=0.42\columnwidth]{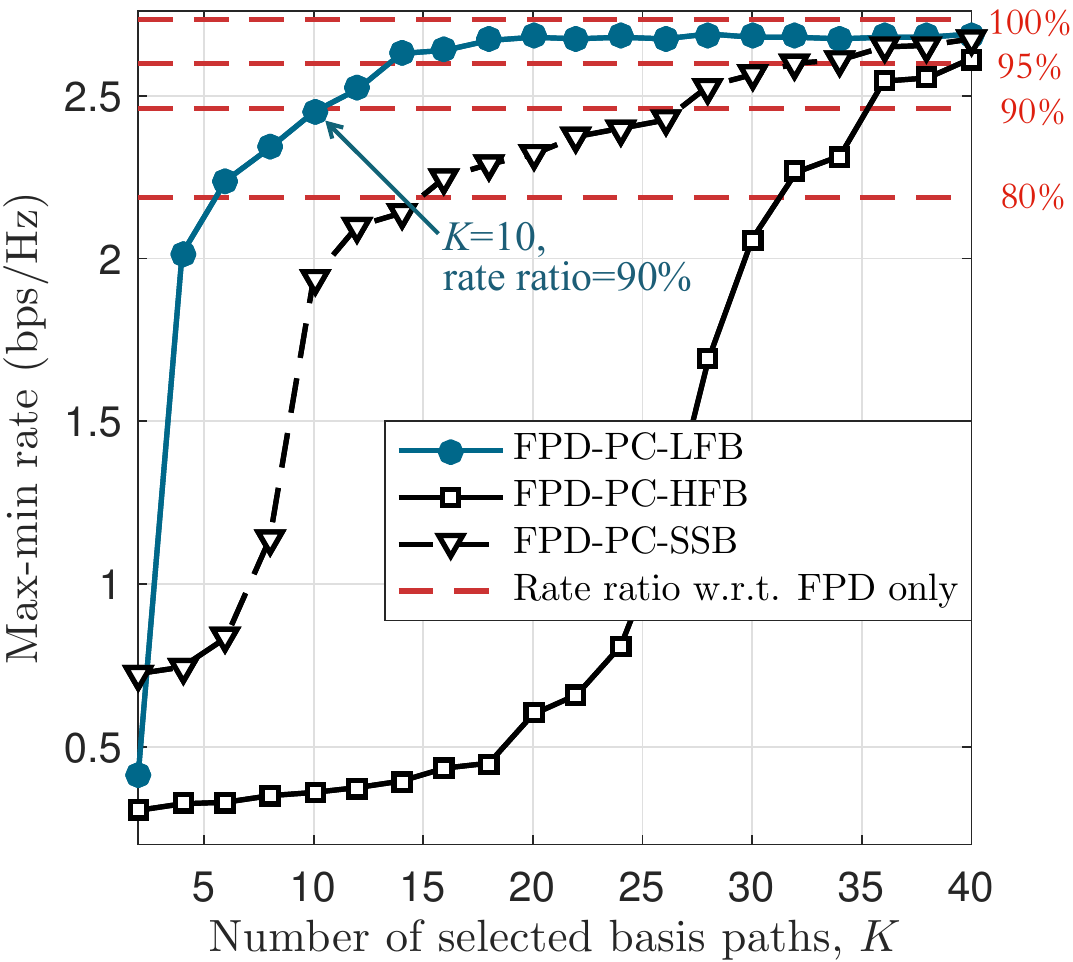} 
	}
	\hspace{15mm}
	\subfigure[Running time versus number of selected basis paths.] 
	{\label{FigB_complex_K}
	\includegraphics[width=0.42\columnwidth]{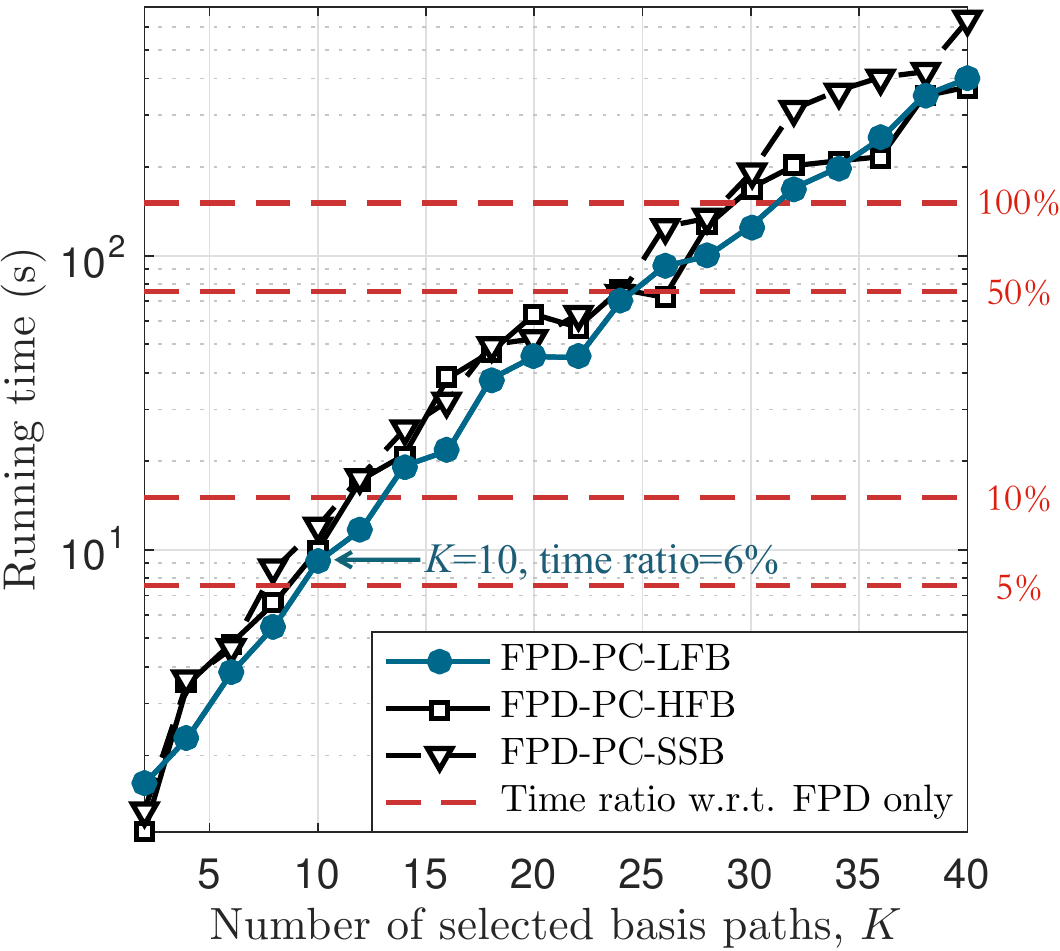}   
	}
	\caption{Effects of parameter $K$ on the performance of the proposed PC-FPD schemes.}     
	\end{figure}
	In Figs.~\ref{FigB_obj_K} and \ref{FigB_complex_K}, we compare the max-min rate and algorithm running time of the proposed  FPD-PC scheme based on the first $K$ (lowest-frequency) Fourier basis paths (named as FPD-PC-LFB) versus the number of selected basis paths ($K$) against the following benchmark schemes: 1) FPD-PC-HFB: the FPD-PC scheme based on the last $K$ (highest-frequency) Fourier basis paths; 2) FPD-PC-SSB: the FPD-PC scheme based on the first $K$ shifted-sine basis paths; 3) FPD only scheme (without PC). We consider $N_{\rm FPD}=200$ and $J=5$ for all the schemes and have the following main observations. First, as $K$ increases,  the max-min rate for the FPD-PC-LFB and FPD-PC-SSB schemes firstly increases rapidly with $K$ and then saturates when $K$ becomes large; while their running time monotonically increases with $K$. The diminishing rate improvement can be explained by the similar reason given for Fig.~\ref{FigA_obj_N}.
	  Second, the proposed FPD-PC-LFB scheme significantly outperforms the FPD-PC-HFB and FPD-PC-SSB schemes as it captures the main path features with low-frequency Fourier basis paths (see detailed reasons in the discussions for Fig.~\ref{PCFitExampleResult} in Section~\ref{SecPathComp}). 
	  Thirdly, with small $K$ (e.g., $K=10$), the proposed FPD-PC-LFB scheme attains nearly $90\%$ of the max-min rate of the FPD only scheme (see Fig.~\ref{FigB_obj_K}), while at the same time, it reduces about $94\%$ of algorithm running time of the FPD only scheme (see Fig.~\ref{FigB_complex_K}). This indicates that we can set small $K$ for the proposed FPD-PC-LFB scheme in practice to substantially  reduce the computational complexity  of the FPD only scheme without scarifying rate performance notably. In addition, it is worth mentioning that given a relatively large $K$ (e.g., $K\ge30$), the proposed FPD-PC-LFB scheme incurs even long running time than the FPD only scheme, since the former scheme involves matrix operation and thus suffers higher computational complexity when $K$ is too large.

	\begin{figure} \centering 
	\includegraphics[width=1\columnwidth]{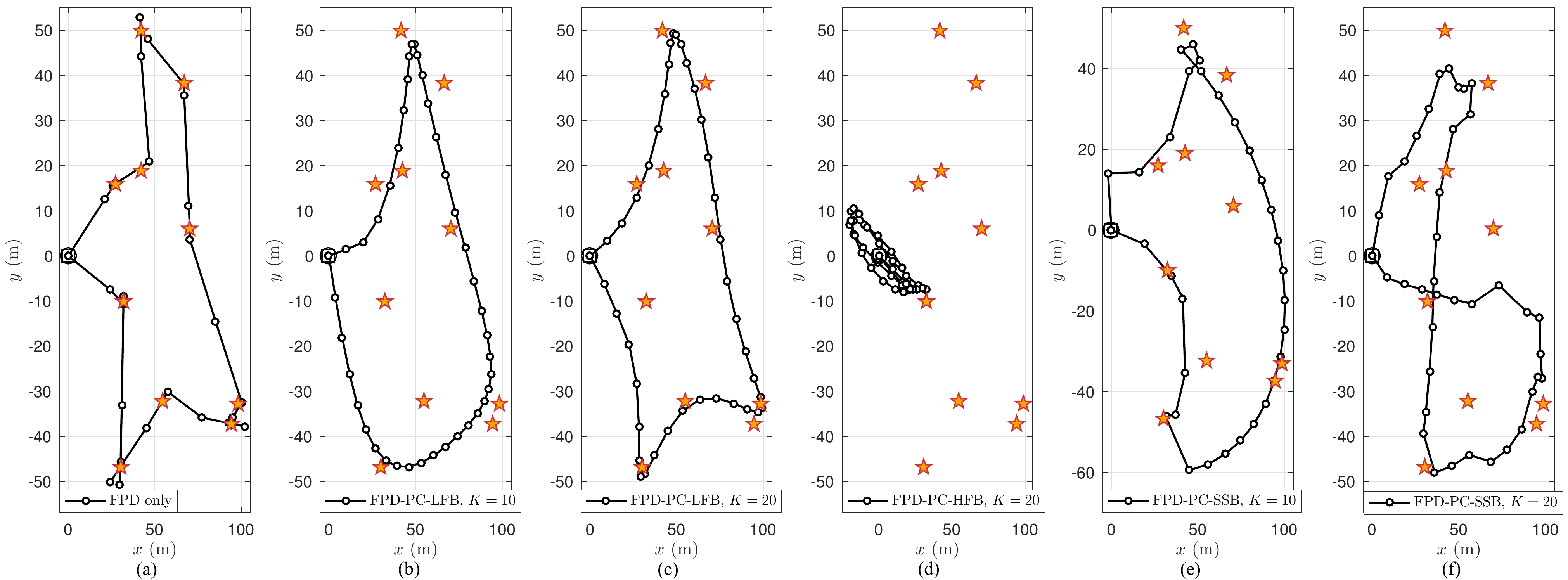} 
	\caption{Optimized UAV trajectories by different  FPD-PC schemes.}     
	\label{FigB_optTrjs}
	\end{figure}

	Fig. \ref{FigB_optTrjs} plots the optimized UAV trajectories by  the FPD-PC scheme under different $K$ and different basis paths as well as that by the FPD only scheme. All the UAV trajectories consist of $N_{\rm FPD}=41$ designable waypoints. Comparing Figs.~\ref{FigB_optTrjs}(a)--\ref{FigB_optTrjs}(c), one can observe that the optimized trajectory by the proposed FPD-PC-LFB scheme approaches to that of the FPD only scheme (without PC) more closely when $K$ is larger  due to more basis paths selected. Next, it is observed in Fig.~\ref{FigB_optTrjs}(d) that different from the FPD-PC-LFB scheme for which the UAV flies nearby the SNs for data collection, the UAV by the FPD-PC-HFB scheme, instead, sticks around the start location even given a relatively large $K$ (e.g., $K=20$). This is because  high-frequency Fourier basis paths render the UAV trajectory with extremely fast path variation and thus can hardly constitute a smooth UAV path.
Besides, given $K=10$ or $K=20$, the optimized UAV trajectories by the FPD-PC-SSB scheme shown in Figs.~\ref{FigB_optTrjs}(e) and \ref{FigB_optTrjs}(f), respectively, are far from optimal  as compared to that in Fig.~\ref{FigB_optTrjs}(a).

\begin{figure}[t] \centering 
	\includegraphics[width=0.48\columnwidth]{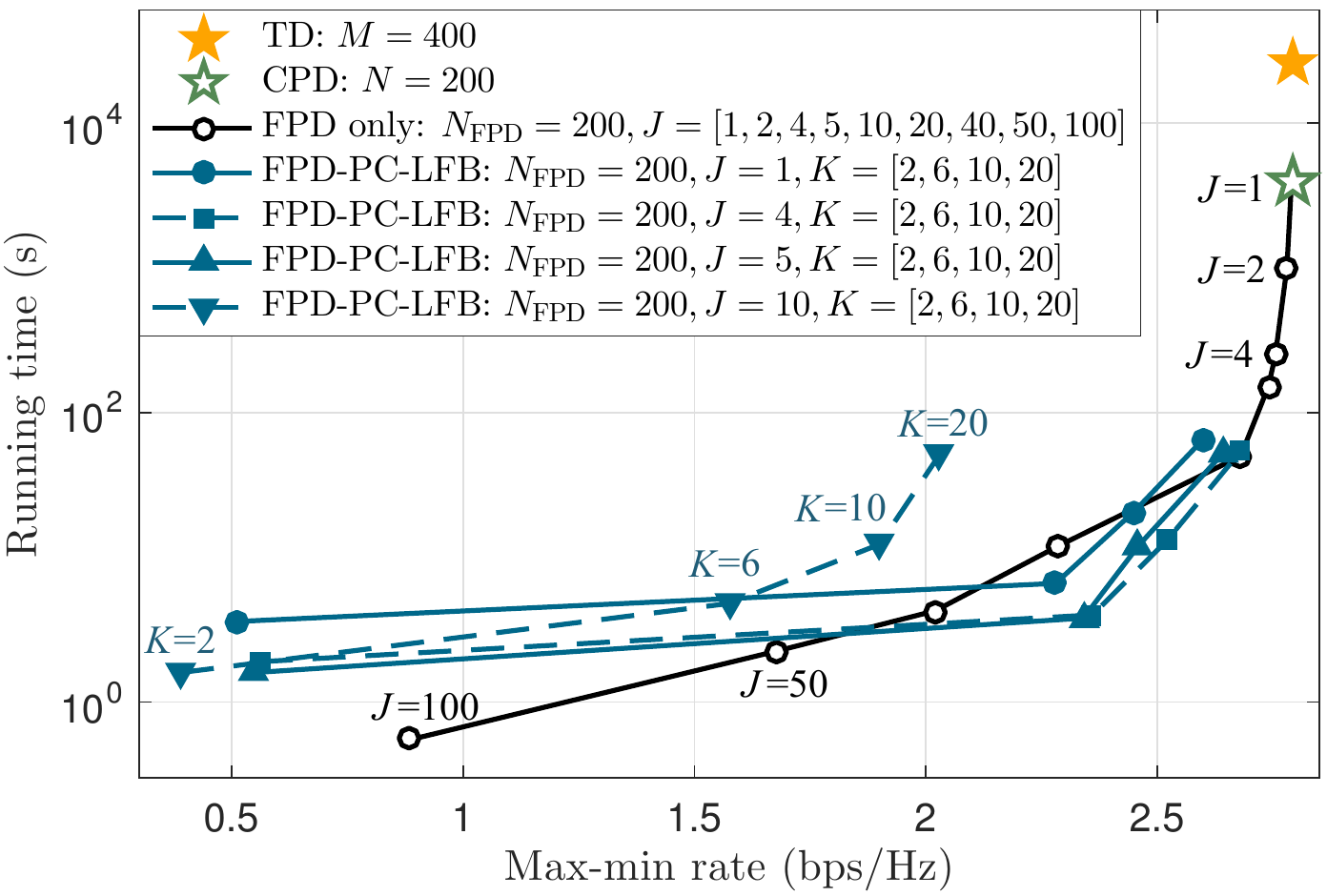} 
	\caption{The trade-off between rate performance and computational complexity for the proposed FPD-PC-LFB scheme.}     
	\label{FigB_Tradeoff}
	\end{figure}

	Last, we show in Fig.~\ref{FigB_Tradeoff} the trade-off between rate performance and computational complexity for the proposed FPD-PC-LFB scheme by jointly adjusting the parameters $J$ and $K$. The TD, CPD, and FPD only schemes are considered as benchmarks for  comparison. 
It is observed that the combination of parameters $\{J, K\}$ significantly affects the performance of the proposed FPD-FC-LFB scheme. On one hand, given the same $K$, it is necessary to properly set a modest value for $J$ (e.g., $J=4$) to achieve good rate performance although the schemes with different values of $J$ incur comparable running time. This is because  setting too small $J$ (e.g., $J=1$) results in a large number of designable waypoints in the UAV path, which cannot be well approximated by the fixed number of basis paths, thus leading to considerable rate performance loss; while setting too large $J$ (e.g., $J=10$) significantly limits the DoF in the UAV trajectory design (even without PC) since only a small  number of designable waypoints are optimized. On the other hand, given a properly chosen $J$ (e.g., $J=4$), it is also necessary to set a proper $K$ to balance the rate performance and design complexity trade-off. Similar observations can be made in Figs.~\ref{FigB_obj_K} and \ref{FigB_complex_K} and thus are omitted for brevity. 
	
\vspace{-5pt}

\section{Conclusions}\label{SecConclusion}
In this paper, we proposed a new and general framework to reduce the computational complexity for the UAV trajectory and communication co-design over the existing TD and CPD schemes. Specifically, to reduce the number of  waypoints to be optimized with CPD, we proposed an FPD scheme that optimizes only  some of the waypoints (called \emph{designable} waypoints) along the path for reducing the trajectory design complexity in the time domain, while all the waypoints (including both designable and non-designable ones) are used in calculating the approximated communication utility along the trajectory for ensuring high  trajectory discretization  accuracy. Moreover, given any  number of designable waypoints, we proposed a novel PC scheme to further  reduce the number of path design variables representing  the designable waypoints by properly selecting a set of basis paths and approximating the path by a superposition  of selected basis paths with optimized path coefficients. Numerical results showed that the proposed FPD and PC schemes  significant reduce the UAV trajectory design complexity yet achieve favorable  rate performance as compared to TD and CPD schemes. The proposed framework is general and can be applied to design UAV trajectories under different channel models and for different purposes.

\vspace{-5pt}
\appendix
\subsection{Proof of Lemma~\ref{lemErrorbound}} \label{Errorbound}
Let  $\Delta u_n$ denote the finite-sum approximation error for the utility function in segment $n$. Then we have
\begin{align}
	\Delta u_n &= \int_{t=0}^{t_n}u(\mathbf{q}_{n-1}+t\mathbf{v}_n) dt-t_n u(\mathbf{q}_n)=\int_{t=0}^{t_n}\left[u(\mathbf{q}_{n-1}+t\mathbf{v}_n)-u(\mathbf{q}_n) \right] dt\nn\\
	&\overset{(a_1)}{=}\int_{t=0}^{t_n}\nabla u(\mathbf{q}_{n-1}+\tilde{\mathbf{q}}_n) (t_n-t)\mathbf{v}_n dt\overset{(a_2)}{\leq} \int_{t=0}^{t_n} |\nabla u(\mathbf{q}_{n-1}+\tilde{\mathbf{q}}_n) \mathbf{v}_n | (t_n-t) dt\nn\\
	&\overset{(a_3)}{\leq} \int_{t=0}^{t_n} D_{u,n} \|\mathbf{v}_n\| (t_n-t) dt =\frac{1}{2} D_{u,n} \|\mathbf{v}_n\| t_n^2=\frac{1}{2} D_{u,n} \Delta_n t_n,
\end{align}
where  $\tilde{\mathbf{q}}_n=t_0\mathbf{v}_n$ with  $t_0\in[0,t_n]$,
$\nabla u$ denotes the gradient of function $u(\cdot)$ w.r.t. $\mathbf{q}$,
$\Delta_n=\|\mathbf{v}_n\| t_n$ is the length of segment $n$, $D_{u,n}= \mathop{\max}_{\mathbf{q}=\mathbf{q}_{n-1}+t_0\mathbf{v}_n, t_0\in[0,t_n]}\|\nabla u(\mathbf{q})\|$.
Note that $(a_1)$ follows from the Rolle mean value theorem, the equality in $(a_2)$ holds when $\nabla u(\mathbf{q}_{n-1}+\tilde{\mathbf{q}}_n)\mathbf{v}_n\ge0$, and the equality in $(a_3)$ holds when $\nabla u(\mathbf{q}_{n-1}+\tilde{\mathbf{q}}_n)$ is constant over segment $n$ and the vectors of $\nabla u$ and $ \mathbf{v}_n$ align in the same direction.
Thus, the finite-sum approximation error over the entire UAV trajectory ${\bf q}(t)$ is
\begin{align}
	E_U=| U({\bf q}(t))- \bar{U}_{{\rm CPD}}(\{{\bf q}_n\},\{t_n\})|&= \l|\sum_{n=1}^N \Delta u_n\r|  \leq \sum_{n=1}^N \frac{1}{2} D_{u,n} \Delta_n t_n \leq  \frac{1}{2} D_{u} \Delta_{\max}^U T,
\end{align}
where $D_{u}=\mathop{\max}_n D_{u,n}$, thus completing the proof.
\subsection{Proof for Example~\ref{Example1}} \label{rateBound}
Consider the achievable rate for the UAV at  the location $\mathbf{q}=[q_{\rm x},q_{\rm y}, q_{\rm z}]$, given as 
\begin{align}
	u(\mathbf{q}) =\log_2\left(1+\frac{P\beta_0}{\|\mathbf{q}-\mathbf{w}\|^2\sigma^2}\right)=\log_2\left(1+\frac{P\beta_0}{\left[(q_{\rm x}-w_{\rm x})^2+(q_{\rm y}-w_{\rm y})^2+q_{\rm z}^2\right]\sigma^2}\right). \label{rateFunctionScale}
\end{align}
With $q_{\rm z}\geq H_{\min}$, it can be shown that  $\|\nabla u(\mathbf{q})\|$ takes its maximum value when $q_{\rm z}=H_{\min}$ and $(q_{\rm x}-w_{\rm x})^2+(q_{\rm y}-w_{\rm z})^2=c_1^2,$
where $c_1^2=\frac{-(2H_{\min}^2+c_2)+\sqrt{(16H_{\min}^4+16c_2H_{\min}^2+c_2^2)}}{6}$ and $c_2=\frac{P\beta_0}{\sigma^2}$. Thus, 
we have $D_u=\mathop{\max}_{\tilde{\mathbf{q}}\in{\bf q}(t)}\|\nabla u(\tilde{\mathbf{q}})\| =\frac{2c_2}{\ln2}\frac{c_1}{(c_1^2+H_{\min}^2)(c_1^2+H_{\min}^2+c_2)}$, thus completing the proof.

\bibliographystyle{IEEEtran}
\bibliography{ref.bib}

\begin{thebibliography}{10}
\providecommand{\url}[1]{#1}
\csname url@samestyle\endcsname
\providecommand{\newblock}{\relax}
\providecommand{\bibinfo}[2]{#2}
\providecommand{\BIBentrySTDinterwordspacing}{\spaceskip=0pt\relax}
\providecommand{\BIBentryALTinterwordstretchfactor}{4}
\providecommand{\BIBentryALTinterwordspacing}{\spaceskip=\fontdimen2\font plus
\BIBentryALTinterwordstretchfactor\fontdimen3\font minus
  \fontdimen4\font\relax}
\providecommand{\BIBforeignlanguage}[2]{{%
\expandafter\ifx\csname l@#1\endcsname\relax
\typeout{** WARNING: IEEEtran.bst: No hyphenation pattern has been}%
\typeout{** loaded for the language `#1'. Using the pattern for}%
\typeout{** the default language instead.}%
\else
\language=\csname l@#1\endcsname
\fi
#2}}
\providecommand{\BIBdecl}{\relax}
\BIBdecl

\bibitem{tutor}
Y.~{Zeng}, Q.~{Wu}, and R.~{Zhang}, ``Accessing from the sky: {A} tutorial on
  {UAV} communications for {5G} and beyond,'' \emph{Proc. IEEE}, vol. 107,
  no.~12, pp. 2327--2375, Dec. 2019.

\bibitem{zeng2016throughput}
Y.~Zeng, R.~Zhang, and T.~J. Lim, ``Throughput maximization for {UAV}-enabled
  mobile relaying systems,'' \emph{IEEE Trans. Commun.}, vol.~64, no.~12, pp.
  4983--4996, Dec. 2016.

\bibitem{chen2018local}
J.~Chen and D.~Gesbert, ``Efficient local map search algorithms for the
  placement of flying relays,'' \emph{IEEE Trans. Wireless Commun.}, vol.~19,
  no.~2, pp. 1305--1319, Feb. 2020.

\bibitem{kang20203d}
Z.~Kang, C.~You, and R.~Zhang, ``{3D} placement for multi-{UAV} relaying: {An}
  iterative {Gibbs}-sampling and block coordinate descent optimization
  approach,'' \emph{arXiv preprint arXiv:2006.09658}, 2020.

\bibitem{chen2017optimum}
Y.~Chen, W.~Feng, and G.~Zheng, ``Optimum placement of {UAV} as relays,''
  \emph{IEEE Commun. Lett.}, vol.~22, no.~2, pp. 248--251, Feb. 2017.

\bibitem{zhan2018energy}
C.~Zhan, Y.~Zeng, and R.~Zhang, ``Energy-efficient data collection in {UAV}
  enabled wireless sensor network,'' \emph{IEEE Wireless Commmu. Lett.},
  vol.~7, no.~3, pp. 328--331, Jun. 2018.

\bibitem{b5}
Q.~{Wu}, Y.~{Zeng}, and R.~{Zhang}, ``Joint trajectory and communication design
  for multi-{UAV} enabled wireless networks,'' \emph{{IEEE} Trans. Wireless
  Commun.}, vol.~17, no.~3, pp. 2109--2121, Mar. 2018.

\bibitem{wu2018capacity}
Q.~Wu, J.~Xu, and R.~Zhang, ``Capacity characterization of {UAV}-enabled
  two-user broadcast channel,'' \emph{IEEE J. Sel. Areas Commun.}, vol.~36,
  no.~9, pp. 1955--1971, Sep. 2018.

\bibitem{gong2018flight}
J.~Gong, T.-H. Chang, C.~Shen, and X.~Chen, ``Flight time minimization of {UAV}
  for data collection over wireless sensor networks,'' \emph{IEEE J. Sel. Areas
  Commun.}, vol.~36, no.~9, pp. 1942--1954, Sep. 2018.

\bibitem{you20193d}
C.~{You} and R.~{Zhang}, ``{3D} trajectory optimization in {Rician} fading for
  {UAV}-enabled data harvesting,'' \emph{{IEEE} Trans. Wireless Commun.},
  vol.~18, no.~6, pp. 3192--3207, Jun. 2019.

\bibitem{you2020hybrid}
C.~You and R.~Zhang, ``Hybrid offline-online design for {UAV}-enabled data
  harvesting in probabilistic {LoS} channel,'' \emph{IEEE Trans. Wireless
  Commun.}, vol.~13, no.~6, pp. 3753--3768, Mar. 2020.

\bibitem{xu2018uav}
J.~Xu, Y.~Zeng, and R.~Zhang, ``{UAV}-enabled wireless power transfer:
  {Trajectory} design and energy optimization,'' \emph{IEEE Trans. Wireless
  Commun.}, vol.~17, no.~8, pp. 5092--5106, Aug. 2018.

\bibitem{hu2019optimal}
Y.~Hu, X.~Yuan, J.~Xu, and A.~Schmeink, ``Optimal {1D} trajectory design for
  {UAV}-enabled multiuser wireless power transfer,'' \emph{IEEE Trans.
  Commun.}, vol.~67, no.~8, pp. 5674--5688, Aug. 2019.

\bibitem{zeng2018cellular}
Y.~Zeng, J.~Lyu, and R.~Zhang, ``Cellular-connected {UAV: Potential},
  challenges, and promising technologies,'' \emph{IEEE Wireless Commun.},
  vol.~26, no.~1, pp. 120--127, Feb. 2018.

\bibitem{zhang2018cellular}
S.~Zhang, Y.~Zeng, and R.~Zhang, ``Cellular-enabled {UAV} communication: {A}
  connectivity-constrained trajectory optimization perspective,'' \emph{IEEE
  Trans. Commun.}, vol.~67, no.~3, pp. 2580--2604, Mar. 2018.

\bibitem{mei2019uplink}
W.~Mei and R.~Zhang, ``Uplink cooperative {NOMA} for cellular-connected
  {UAV},'' \emph{IEEE J. Sel. Topics Signal Process.}, vol.~13, no.~3, pp.
  644--656, Jun. 2019.

\bibitem{hua2018power}
M.~Hua, Y.~Wang, Z.~Zhang, C.~Li, Y.~Huang, and L.~Yang, ``Power-efficient
  communication in {UAV}-aided wireless sensor networks,'' \emph{IEEE Commun.
  Lett.}, vol.~22, no.~6, pp. 1264--1267, June 2018.

\bibitem{zhou2018computation}
F.~Zhou, Y.~Wu, R.~Q. Hu, and Y.~Qian, ``Computation rate maximization in
  {UAV}-enabled wireless-powered mobile-edge computing systems,'' \emph{IEEE J.
  Sel. Areas Commun.}, vol.~36, no.~9, pp. 1927--1941, Sep. 2018.

\bibitem{mozaffari2016optimal}
M.~Mozaffari, W.~Saad, M.~Bennis, and M.~Debbah, ``Optimal transport theory for
  power-efficient deployment of unmanned aerial vehicles,'' in \emph{Proc. IEEE
  Intl. Conf. Commun. (ICC)}, May 2016, pp. 1--6.

\bibitem{mozaffari2015drone}
M.~{Mozaffari}, W.~{Saad}, M.~{Bennis}, and M.~{Debbah}, ``Drone small cells in
  the clouds: Design, deployment and performance analysis,'' in \emph{Proc.
  {IEEE} Global Commun. Conf. (Globecom)}, Dec. 2015, pp. 1--6.

\bibitem{b6}
J.~{Lyu}, Y.~{Zeng}, R.~{Zhang}, and T.~J. {Lim}, ``Placement optimization of
  {UAV}-mounted mobile base stations,'' \emph{{IEEE} Commun. Lett.}, vol.~21,
  no.~3, pp. 604--607, Mar. 2017.

\bibitem{alzenad20173}
M.~Alzenad, A.~El-Keyi, and H.~Yanikomeroglu, ``3-{D} placement of an unmanned
  aerial vehicle base station for maximum coverage of users with different
  {QoS} requirements,'' \emph{IEEE Wireless Commun. Lett.}, vol.~7, no.~1, pp.
  38--41, Feb. 2018.

\bibitem{zhang2019receding}
J.~Zhang, Y.~Zeng, and R.~Zhang, ``Receding horizon optimization for
  energy-efficient {UAV} communication,'' \emph{IEEE Wireless Commun. Lett.},
  vol.~9, no.~4, pp. 490--494, Apr. 2019.

\bibitem{lee2019uav}
J.-H. Lee, K.-H. Park, Y.-C. Ko, and M.-S. Alouini, ``A {UAV}-mounted free
  space optical communication: {Trajectory} optimization for flight time,''
  \emph{IEEE Trans. Wireless Commun.}, vol.~19, no.~3, pp. 1610--1621, Mar.
  2020.

\bibitem{guo2020novel}
Y.~Guo, S.~Yin, J.~Hao, and Y.~Du, ``A novel trajectory design approach for
  {UAV} based on finite {Fourier} series,'' \emph{IEEE Wireless Commun. Lett.},
  vol.~9, no.~5, pp. 671--674, May 2020.

\bibitem{shen2020multi}
C.~Shen, T.-H. Chang, J.~Gong, Y.~Zeng, and R.~Zhang, ``Multi-{UAV}
  interference coordination via joint trajectory and power control,''
  \emph{IEEE Trans. Signal Process.}, vol.~68, pp. 843--858, Jan. 2020.

\bibitem{xu2020low}
K.~Xu, M.-M. Zhao, Y.~Cai, and L.~Hanzo, ``Low-complexity joint power
  allocation and trajectory design for {UAV}-enabled secure communications with
  power splitting,'' \emph{arXiv preprint arXiv:2008.10015}, 2020.

\bibitem{ben2001lectures}
A.~Ben-Tal and A.~Nemirovski, \emph{Lectures on modern convex optimization:
  {Analysis}, algorithms, and engineering applications}.\hskip 1em plus 0.5em
  minus 0.4em\relax Siam, 2001, vol.~2.

\bibitem{grant2008cvx}
M.~Grant, S.~Boyd, and Y.~Ye, ``{CVX}: Matlab software for disciplined convex
  programming,'' 2008.

\end{thebibliography}

\end{document}